\documentclass[openacc]{rsproca_new}

\newtheorem{theorem}{\bf Theorem}[section]

\usepackage{stmaryrd}
\usepackage{ulem}
\usepackage{mathtools}
\usepackage{appendix}

\usepackage[linesnumbered,ruled,vlined]{algorithm2e}

\SetCommentSty{mycommfont}
\SetKwInput{KwInput}{Input}                
\SetKwInput{KwOutput}{Output}  

\makeatletter
\def\widebreve#1{\mathop{\vbox{\m@th\ialign{##\crcr\noalign{\kern3\p@}%
      \brevefill\crcr\noalign{\kern3\p@\nointerlineskip}%
      $\hfil\displaystyle{#1}\hfil$\crcr}}}\limits}

\def\brevefill{$\m@th \setbox\z@\hbox{$\braceld$}%
  \bracelu\leaders\vrule \@height\ht\z@ \@depth\z@\hfill\braceru$}
\makeatletter

\begin{document}

\definecolor{forestgreen}{RGB}{1, 93, 45}
\definecolor{pigmentgreen}{RGB}{0, 165, 80}
\definecolor{custompurple}{RGB}{129, 19, 239}
\definecolor{customred}{RGB}{208, 2, 27}
\definecolor{customyelloworange}{RGB}{255, 137, 0}
\definecolor{customturquoise}{RGB}{7, 143, 181}
\definecolor{palegreen}{rgb}{0.263,0.804,0.502}

\newcommand{\matcomment}[1]{\textcolor{blue}{\bf{[Mathieu: #1]}}}
\newcommand{\jeffcomment}[1]{\textcolor{forestgreen}{\bf{[Jeff: #1]}}}
\newcommand{\ricomment}[1]{\textcolor{customred}{\bf{[Ricardo: #1]}}}
\newcommand{\youcomment}[1]{\textcolor{customturquoise}{\bf{[Youssef: #1]}}}

\newcommand{\diag}[1]{\text{diag}\left(#1\right)}
\newcommand{\epde}{ePDE}
\newcommand{\pde}{PDE}
\newcommand{\psd}{PSD}

\newcommand{\jeff}[1]{\textcolor{forestgreen}{#1}}
\newcommand{\ric}[1]{\textcolor{customred}{#1}}
\newcommand{\mat}[1]{\textcolor{blue}{#1}}
\newcommand{\youssef}[1]{\textcolor{customturquoise}{#1}}

\newcommand{\eqn}{equation}
\newcommand{\Eqn}{Equation}

\newcommand{\manifold}{\Omega}
\newcommand{\linop}{\mathcal{L}}
\newcommand{\greenf}[1]{G_{#1}}

\newcommand{\argmin}{argmin\;}
\newcommand{\argmax}{argmax\;}

\newcommand{\atm}{\texttt{AdaptiveTransportMap.jl}}

\newcommand{\eg}{e.g.\;}
\newcommand{\ie}{i.e.\;}

\newcommand{\BB}{\boldsymbol}
\newcommand{\be}{\begin{equation}}
\newcommand{\ee}{\end{equation}}

\newcommand{\ba}{\begin{align}}
\newcommand{\ea}{\end{align}}

\newcommand{\real}[1]{\mathbb{R}^{#1}}
\newcommand{\complex}[1]{\mathbb{C}^{#1}}

\newcommand{\id}[1]{\BB{I}_{#1}}
\newcommand{\zero}[1]{\BB{0}_{#1}}
\newcommand{\one}{\BB{1}}

\definecolor{forecast}{RGB}{19, 24, 143}
\definecolor{analysis}{RGB}{208, 2, 27}

\definecolor{posterior}{rgb}{0.263,0.804,0.502}
\definecolor{truth}{rgb}{0.125,0.698,0.667}
\definecolor{enkf}{rgb}{0.933,0.251,0.0}

\definecolor{smap0}{rgb}{0.686, 0.624, 0.11}
\definecolor{smap1}{rgb}{0.275,0.51,0.706}  
\definecolor{smap2}{rgb}{0.545, 0.0, 0.0}

\definecolor{ccfd}{rgb}{0.0,0.0,0.502}
\definecolor{cenkf}{rgb}{0.0,0.502,0.502}
\definecolor{cetkf}{rgb}{0.933,0.251,0.0}  
\definecolor{cetkfconf}{RGB}{246, 157, 127}
\definecolor{cgust}{rgb}{0.196,0.804,0.196}
\definecolor{ctmap}{rgb}{0.502,0.0,0.502}
\definecolor{cgold}{rgb}{0.8, 0.61, 0.11}

\newcommand{\commentcode}[1]{\textcolor{cenkf}{\footnotesize{\texttt{\% #1}}}}

\definecolor{customblue}{RGB}{19, 24, 143}

\definecolor{orangered}{RGB}{168,58,8}
\definecolor{customyellow}{RGB}{252,227,3}
\definecolor{customgreen}{RGB}{0,158,115}

\newcommand\crule[3][black]{\textcolor{#1}{\rule{#2}{#3}}}
\newcommand\lcolor[1]{\crule[#1]{2mm}{2mm}}
\newcommand\squarecolor[1]{\crule[#1]{2mm}{2mm}}
\newcommand\linecolor[1]{\textcolor{#1}{\rule[1pt]{4mm}{0.8mm}}}

\newcommand{\lesp}{LESP}
\newcommand{\lespc}{$\mbox{LESPc}$}
\newcommand{\cn}{$C_n$}
\newcommand{\ts}{t^\star}


\newcommand{\statef}{{\state^f}}
\newcommand{\statefi}{{\state^{f,i}}}
\newcommand{\statea}{{\state^a}}
\newcommand{\stateai}{{\state^{a,i}}}
\newcommand{\statebar}{\overline{\state}}
\newcommand{\statebara}{\statebar^a}
\newcommand{\statebarf}{\statebar^f}

\newcommand{\covm}{\BB{Q}}
\newcommand{\covo}{\BB{R}}
\newcommand{\score}{\BB{\Omega}}

\newcommand{\Ne}{N_e}
\newcommand{\Nx}{N_x}
\newcommand{\Ny}{N_y}

\newcommand{\K}{\BB{K}}

\newcommand{\KH}{\BB{KH}}

\newcommand{\Pa}{\BB{P}^a}
\newcommand{\Pf}{\BB{P}^f}

\newcommand{\Pbar}{\overline{\BB{P}}}
\newcommand{\Pbara}{\overline{\BB{P}}^a}
\newcommand{\Pbarf}{\overline{\BB{P}}^f}


\newcommand{\Ens}{\BB{E}}
\newcommand{\Ensa}{\BB{E}^a}
\newcommand{\Ensf}{\BB{E}^f}

\newcommand{\ensemble}[1]{\left\{#1\right\}}

\newcommand{\X}{\BB{X}}
\newcommand{\Xa}{\X^a}
\newcommand{\Xf}{\X^f}
\newcommand{\Xp}{{\BB{X}^{\prime}}}
\newcommand{\Xpa}{{\BB{X}^{\prime a}}}
\newcommand{\Xpf}{{\BB{X}^{\prime f}}}
\newcommand{\Xpai}{{\BB{X}^{\prime a, i}}}
\newcommand{\Xpfi}{{\BB{X}^{\prime f, i}}}
\newcommand{\Xbar}{\overline{\BB{X}}}
\newcommand{\Xbara}{\Xbar^a}
\newcommand{\Xbarf}{\Xbar^f}

\newcommand{\Ano}{\BB{A}}
\newcommand{\Anoa}{\Ano^a}
\newcommand{\Anof}{\Ano^f}

\newcommand{\covobsp}{{\BB{W}^{\prime}}}
\newcommand{\covobspi}{\BB{W}^{\prime, i}}

\newcommand{\Ypf}{{\BB{Y}^{\prime f}}}
\newcommand{\Ypfi}{\BB{Y}^{\prime f, i}}

\newcommand{\ybarf}{\overline{\BB{y}}^f}

\newcommand{\U}{\BB{U}}
\newcommand{\J}{\BB{J}}
\newcommand{\V}{\BB{V}}
\newcommand{\T}{\BB{T}}
\newcommand{\G}{\BB{G}}
\newcommand{\sqG}{{\BB{G}^{1/2}}}

\newcommand{\Rho}{\BB{\rho}}


\newcommand\given[1][]{\:#1\vert\:}

\newcommand{\N}{\mathcal{N}}

\newcommand{\E}[2]{\mathrm{E}_{#1}\left[#2\right]}
\newcommand{\cov}[1]{\BB{\Sigma}_{#1}}
\newcommand{\scov}[1]{\widehat{\BB{\Sigma}}_{#1}}
\newcommand{\mean}[1]{\BB{\mu}_{#1}}
\newcommand{\smean}[1]{\widehat{\mu}_{#1}}
\newcommand{\dmat}[1]{\BB{d}_{#1}}

\newcommand{\enkf}{EnKF}
\newcommand{\senkf}{sEnKF}
\newcommand{\lrenkf}{LREnKF}
\newcommand{\locenkf}{LocEnKF}
\newcommand{\etkf}{ETKF}
\newcommand{\smf}{SMF}

\newcommand{\state}{\BB{x}}
\newcommand{\meas}{\BB{y}}
\newcommand{\measbarf}{\overline{\meas}^f}
\newcommand{\dyn}{\BB{f}}
\newcommand{\obs}{\BB{h}}
\newcommand{\Obs}{\BB{H}}
\newcommand{\tObs}{\tilde{\BB{H}}}
\newcommand{\Noisedyn}{\BB{\mathsf{W}}}
\newcommand{\Noiseobs}{\BB{\mathcal{E}}}
\newcommand{\noisedyn}{\BB{w}}
\newcommand{\noiseobs}{{\BB{\epsilon}}}
\newcommand{\covdyn}{\BB{W}}
\newcommand{\covobs}{\BB{V}}

\newcommand{\pdf}[1]{\pi_{#1}}
\newcommand{\pdfprior}{\pdf{\State}}
\newcommand{\pdflik}{\pdf{\Meas \given \State}}
\newcommand{\pdfjoint}{\pdf{\Meas, \State}}
\newcommand{\pdfpost}{\pdf{\State \given \Meas}}

\newcommand{\State}{\BB{\mathsf{X}}}
\newcommand{\Meas}{\BB{\mathsf{Y}}}

\newcommand{\Stateup}{\boldsymbol{\mathcal{X}}}
\newcommand{\Measup}{\boldsymbol{\mathcal{Y}}}

\newcommand{\var}{\BB{z}}
\newcommand{\Var}{\BB{\mathsf{Z}}}

\newcommand{\proj}{\BB{P}}

\newcommand{\tmap}{\BB{T}}
\newcommand{\tmappush}{\BB{T}_\sharp}
\newcommand{\tmappull}{\BB{T}^\sharp}

\newcommand{\smap}{\BB{S}}
\newcommand{\smappush}{\BB{S}_\sharp}
\newcommand{\smappull}{\BB{S}^\sharp}

\newcommand{\umap}{\BB{U}}

\newcommand{\Hmap}{\BB{\mathcal{H}}}
\newcommand{\Hmapdelta}{\BB{\mathcal{H}}_\Delta}
\newcommand{\Hmapk}{\mathcal{H}_k}
\newcommand{\Hmapkh}{\mathcal{H}_{k,a}}

\newcommand{\ui}{\BB{\mathfrak{u}}_i}
\newcommand{\uk}{\BB{\mathfrak{u}}_k}
\newcommand{\psij}{\psi_j}
\newcommand{\Uk}{U^k}

\newcommand{\Cx}{\BB{C}_{\State}}
\newcommand{\Cy}{\BB{C}_{\Meas}}
\newcommand{\sCx}{\widehat{\BB{C}}_{\State}}
\newcommand{\sCy}{\widehat{\BB{C}}_{\Meas}}
\newcommand{\Sx}{\BB{S}_{\State}}
\newcommand{\Sy}{\BB{S}_{\Meas}}

\newcommand{\rx}{r_{\State}}
\newcommand{\ry}{r_{\Meas}}
\newcommand{\Kbreve}{\breve{\BB{K}}}

\newcommand{\Q}{\BB{Q}}

\def\dkl#1#2{\mathcal{D}_{\mathrm{KL}}\left(#1 || #2\right)}

\newcommand{\rmse}{\mbox{RMSE}}


\newcommand{\recti}[2]{\mathcal{R}_{#1}(#2)}
\newcommand{\lossJ}[2]{\mathcal{J}_{#1}(#2)}
\newcommand{\lossL}[2]{\mathcal{L}_{#1}(#2)}
\newcommand{\coeff}{c_{\BB{\alpha}}}

\newcommand{\sketch}{\BB{\Omega}}
\newcommand{\sketchbis}{\BB{\Gamma}}
\newcommand{\tbi}{\texttt{TransportBasedInference.jl}}
\newcommand{\dist}{distribution}
\newcommand{\kr}{Knothe-Rosenblatt}
\newcommand{\rea}{rearrangement}

\newcommand{\pot}{\phi}
\newcommand{\cpot}{F}
\newcommand{\stream}{\psi}

\newcommand{\ez}{\BB{e}_z}

\newcommand{\position}{\BB{r}}
\newcommand{\positionp}{\BB{r}^\prime}
\newcommand{\zp}{z^\prime}

\newcommand{\vel}{\BB{v}}
\newcommand{\vort}{\BB{\omega}}
\newcommand{\dilat}{\theta}

\newcommand{\diver}[1]{\nabla\cdot#1}
\newcommand{\curl}[1]{\nabla\times#1}

\newcommand{\lap}{\nabla^2}

\newcommand{\ddt}[1]{\frac{\mathrm{d}#1}{\mathrm{d}t}}
\newcommand{\pddt}[1]{\frac{\partial#1}{\partial t}}

\newcommand{\circu}{\Gamma}
\newcommand{\flux}{Q}
\newcommand{\strength}{S}

\newcommand{\Ustream}{\BB{U}_{\infty}}
\newcommand{\Wstream}{W_{\infty}}

\newcommand{\re}[1]{\mathfrak{Re}\left(#1\right)}
\newcommand{\im}[1]{\mathfrak{Im}\left(#1\right)}

\newcommand{\kernel}[1]{\BB{k}\left( #1 \right)}
\newcommand{\kernelblob}[1]{\tilde{\BB{k}}( #1 )}

\newcommand{\Jidx}{\mathsf{J}}
\newcommand{\Kidx}{\mathsf{K}}
\newcommand{\Iidx}{\mathsf{I}}
\newcommand{\Lidx}{\mathsf{L}}

\newcommand{\Nsing}{\mathsf{N}}
\newcommand{\kirch}{Kirchhoff}

\newcommand{\setsource}{\mathbb{S}}
\newcommand{\settarget}{\mathbb{T}}

\newcommand{\chol}[1]{\BB{L}_{#1}}
\newcommand{\vander}[1]{\mbox{Vander}_{#1}}
\newcommand{\dobs}{\nabla_{\var}\obs}

\newcommand{\diff}[2]{\frac{\mathrm{d}#1}{\mathrm{d}#2}}
\newcommand{\diffp}[2]{\frac{\partial#1}{\partial #2}}

\renewcommand{\eg}{e.g., }
\renewcommand{\ie}{i.e., }

\title{A low-rank ensemble Kalman filter for elliptic observations}

\author{
Mathieu Le Provost$^{1}$, Ricardo Baptista$^{2}$, Youssef Marzouk$^{2}$ and Jeff D.\ Eldredge$^{1}$}

\address{$^{1}$Mechanical \& Aerospace Engineering Department,\\
	University of California, Los Angeles, \\
	Los Angeles, CA 90095, USA\\
$^{2}$Department of Aeronautics and Astronautics, 
	Massachusetts Institute of Technology\\
	Cambridge, MA, 02139, USA\\}

\subject{fluid mechanics, mathematical modelling, applied mathematics}


\corres{Mathieu Le Provost\\
\email{mathieuleprovost1@gmail.com}}

\begin{abstract}
We propose a regularization method for ensemble Kalman filtering (EnKF) 
with elliptic observation operators.
Commonly used EnKF regularization methods suppress state correlations at long distances. For observations described by elliptic partial differential equations, such as the pressure Poisson equation (PPE) in incompressible fluid flows, distance localization should be used cautiously, as we cannot disentangle slowly decaying physical interactions from spurious long-range correlations. This is particularly true for the PPE, in which distant vortex elements couple nonlinearly to induce pressure. Instead, these inverse problems have a low effective dimension: low-dimensional projections of the observations strongly inform a low-dimensional subspace of the state space. We derive a low-rank factorization of the Kalman gain based on the spectrum of the Jacobian of the observation operator. The identified eigenvectors generalize the source and target modes of the multipole expansion, independently of the underlying spatial distribution of the problem. Given rapid spectral decay, inference can be performed in the low-dimensional subspace spanned by the dominant eigenvectors. This low-rank EnKF is assessed on dynamical systems with Poisson observation operators, where we seek to estimate the positions and strengths of point singularities over time from potential or pressure observations. We also comment on the broader applicability of this approach to elliptic inverse problems outside the context of filtering.
\end{abstract}


\begin{fmtext}

\end{fmtext}

\maketitle


In data assimilation, observational data from a real dynamical system are used to correct model predictions of the system's state. This paper focuses on the assimilation of elliptic observations---i.e., observations related to the state via an elliptic partial differential equation---in a dynamically evolving process. \textit{Filtering} is a classic problem in data assimilation, where we aim to estimate the state of a dynamical system at time $k$, $\state_k \in \real{n}$, given all observations available up to that time. State-of-the-art results for high-dimensional filtering problems are often obtained with the ensemble Kalman filter (\enkf{})~\cite{evensen1994sequential}. The \enkf{} recursively updates a set of $M$ realizations of the state, called ``particles'' or ``ensemble members'' $\{\state^1, \ldots, \state^M\}$, to form an empirical approximation of the filtering distribution. Within each assimilation cycle, we perform a \textit{forecast step} followed by an \textit{analysis step}. In the forecast step, the filtering ensemble is propagated through the dynamical model for a finite time to generate samples of the forecast distribution. The analysis step updates the forecast ensemble by conditioning on newly available observations of the true system. The \enkf{} performs the analysis step by using the forecast ensemble to form a Monte Carlo approximation of the Kalman gain. For typical applications, a \textit{limited} number of ensemble members ($M \sim 100$) are tracked compared to the dimension of the state variable ($n \sim 10^4-10^9$) and the dimension of the observation ($d \sim 10^2-10^5$)~\cite{evensen2009ensemble}. Hence, the empirical state and observation covariance matrices are rank-deficient, and suffer from both sampling errors and spurious long-range correlations. While in many high-dimensional problems the \enkf{} can successfully track the state with very limited samples,  this success is predicated on an appropriate regularization of the empirical Kalman gain. Predominant regularization techniques assume that the observations are \textit{local}, \ie that an observation only provides information about a subset of the state variables which are close in physical distance. This assumption is supported by the rapid decay of the correlations between state and observation variables. In this setting, distance localization cuts off long-range correlations~\cite{asch2016data}.

In this paper, we are interested in filtering problems where the observations correspond to \textit{non-local} functions of the state, such as integrals of linear and nonlinear functions of the state~\cite{leeuwen2019non}. As examples, we can cite the radiance measured by satellites, heat or mass fluxes through surfaces, forces measured on a body immersed in a fluid---or, as we address this paper, solutions of elliptic partial differential equations (\pde{}s). An elliptic \pde{} is given by
\begin{equation}
\label{eqn:epde}
    \linop \BB{u}_k(\position)  = \mathcal{\BB{q}}_k(\position; \state_k),\ \position \in \manifold,
\end{equation} 
where $\manifold$ is the physical space (a subset of  $\real{2}$ or $\real{3}$), $\position \in \manifold$ denotes the point of evaluation of~\eqref{eqn:epde}, $\linop$ is an elliptic linear operator (\eg the Laplacian $\lap$, in which case~\eqref{eqn:epde} is a Poisson equation), and $\mathcal{\BB{q}}_k$ is a spatially distributed forcing term that in general depends nonlinearly on the state $\state_k$. We will consider filtering problems where, at each assimilation step $k$, we seek to estimate the state $\state_k$ from a small number of pointwise evaluations of the solution $\BB{u}_k$ of the elliptic \pde{}~\eqref{eqn:epde}; in other words, we will observe $\BB{u}_k$ with noise at only a few points $\position \in \manifold$. Up to a homogeneous part, the solution $\BB{u}_k$ can be obtained by convolution of the Green's function of the elliptic operator $\linop$ with the forcing term $\mathcal{\BB{q}}_k$ over $\manifold$. The observations thus inherit the non-locality and the nonlinearity of the Green's function, along with further nonlinear state dependence in the forcing term $\mathcal{\BB{q}}_k$.


For a Poisson equation, the Green's function of the Laplacian decays logarithmically or algebraically (based on the dimension of the physical space) as a function of distance, so we expect long-range physical interactions between the state and observation variables. Distance localization should be used carefully, as it will remove \textit{all} long-range correlations, whether physical or spurious. In this paper, we develop a novel regularization technique to unambiguously assimilate such \textit{non-local} observations by exploiting the known structure of the solutions of \eqref{eqn:epde}. We emphasize that Kalman-based ensemble filters, including the present work,  treat the dynamical model as a black box, and only need to evaluate it at a set of samples in the forecast step. Thus, our methodology is readily applicable to a wide range of filtering problems with elliptic observations such as \cite{gwirtz2021testbed, lang2017data} in electromagnetism  or \cite{colburn2011state, dasilva2020flow, li2022data} in incompressible fluid flows.  

In this paper, we focus our discussion on the representative context of incompressible fluid mechanics. In these problems, the flow field is most compactly represented by the vorticity, or curl of the velocity field. We use a low-dimensional Lagrangian representation of the flow field by tracking the positions and strengths of a small collection of point vortices. Inviscid point vortex models have been a long-standing tool to model and explain incompressible fluid phenomena~\cite{sarpkaya1975inviscid, eldredgebook, leprovost2021ensemble}. In this paper, we focus on filtering problems where we seek to estimate the characteristics of point singularities over time from limited potential or pressure observations. Despite their relative low dimensionality, these problems can be particularly challenging as the transport equation for point vortices (the Biot--Savart law) and the observation model (the pressure Poisson equation) involve the resolution of Poisson equations like \eqref{eqn:epde}, whose forcing terms nonlinearly couple all the singularities' contributions. 


In this work, we introduce a new regularization approach motivated by the low-rank approximation of \textit{inverse problems}. Indeed, given the indirect relationship between state and observations, each assimilation step in our current setting can be viewed as the solution of an inverse problem. It is natural to cast this inverse problem in a Bayesian setting, where the parameters of interest comprise the state $\state_k$, the prior distribution is approximated empirically by the forecast ensemble, and the forward model (and hence the likelihood function) follow from \eqref{eqn:epde}.
In a wide range of inverse problems, observations are informative---relative to the prior---about only a low-dimensional subspace of the parameter space~\cite{cui2014likelihood, spantini2015optimal, zahm2018certified, cui2021data}. There are several reasons for this structure: information about the state is lost through the smoothing action of the forward operator, as well as its nonlinearity; observations may be few in number, relative to the state dimension; and noise in the observations makes it difficult to extract small-amplitude features of the parameters. 
In the linear--Gaussian setting, Spantini et al.~\cite{spantini2015optimal} exploit these ideas by identifying the optimal (in a certain precise sense) prior-to-posterior update of any given dimension in the parameter space, based on an eigensystem that combines the forward operator, the prior covariance matrix, and the noise covariance matrix. Similarly, in the linear--Gaussian setting, Giraldi et al.~\cite{giraldi2018optimal} identify low-dimensional projections of the observations that are optimal in the sense of preserving mutual information with the parameters. A duality between parameter and observation reduction in the linear--Gaussian setting is further elucidated in \cite[Sec.~3.6.2]{jagalur2020batch}.
Cui et al.~\cite{cui2021data} generalize the parameter reduction approach of \cite{spantini2015optimal} to the \textit{nonlinear/non-Gaussian setting}, based on integrals of the Jacobian of the nonlinear forward model. Here, the strict duality between parameter and observation space eigensystems is lost, and observation reduction requires additional effort. Inspired by \cite{cui2021data}, here we propose an analogous method to identify the most informative directions in the observation space, for the nonlinear/non-Gaussian setting.
%
%
%
This companion dimension reduction is motivated by the potential for redundancy and low observability of some sensors, as well as the spatial structure of the data. Ideas for such a \textit{joint} dimension reduction have already been exploited in the \textit{fast multipole method (FMM)}, which accelerates the solution of the Poisson equation by clustering interactions between ``sources'' and ``targets''~\cite{greengard1987fast, ethridge2001new}. Our algorithm provides a systematic method to identify these clusters of variables from the Jacobian of the observation model, without knowledge of the spatial organization of the singularities and the evaluation points. 

We then use this identified structure to \textit{regularize the \enkf}: specifically, we construct a low-rank factorization of the Kalman gain based on the identified modes for the states and observations. We observe that for elliptic observation operators, a few modes can very accurately approximate the row and column spaces of the Kalman gain. 
The analysis step of the proposed observation-informed low-rank EnKF (\lrenkf{}) can be summarized by the following sequence of steps: identify and project the state and observation variables on the leading directions in the state and observation spaces, compute the Kalman gain and apply the linear update of the \enkf{} in these reduced coordinates, and lift the result to the original variable space.  


The rest of the paper is organized as follows. Section~\ref{sec:potentialflow} gives a basic outline of the model equations of potential flow theory, and highlights the nonlinear and elliptic properties of the dynamical and observation models. Section~\ref{sec:enkf} reviews the classical treatment of the EnKF and presents our low-rank ensemble Kalman filter. Example problems are treated in Section~\ref{sec:examples}. Concluding remarks follow in Section~. For the sake of conciseness, certain details about the model equations and the proposed low-rank ensemble Kalman filter, including a pseudocode, are provided in the Supplementary Material.

\section{Model equations of potential flow}

\label{sec:potentialflow}

This section gives a minimal outline of the model equations governing an incompressible flow described by advecting point singularities. Further details of the equations are given in Supplementary Material \ref{secSM:biotsavart}  and can be found in many references, such as \cite{eldredgebook}. Though we focus in this work on two-dimensional fluid mechanics, the equations are representative of problems in many areas of physics that follow the template of equation \eqref{eqn:epde}, in two or three dimensions, and the techniques we develop are easily extendable to these other contexts.

In this study, we use both vector notation and complex notation, the former for emphasizing the generality and the latter for compactly representing some of the basic equations. A point in space will either be referred to in vector notation as $\position =  (x, y) \in \real{2}$, or in complex notation as $z = x + i y$. The conjugate of a complex number $z = \re{z} + i \im{z}$ is denoted as $\overline{z} \equiv \re{z} - i \im{z}$. We denote the canonical basis of $\real{3}$ as $(\BB{e}_1, \BB{e}_2, \BB{e}_3)$.

In general, a vector field $\vel(\position, t)$ such as fluid velocity can be obtained from the sum of the gradient of a scalar potential $\pot(\position, t)$ and the curl of a vector potential (or streamfunction) $\stream(\position, t)$, each of which satisfies a Poisson equation, with right-hand sides given respectively by the divergence of $\vel$ and the (negative) curl of $\vel$, \ie the vorticity. A potential flow refers to a flow field in which those right-hand-side quantities occupy, at most, singular regions of zero volume~\cite{eldredgebook}. We consider a set of $N$ singularities located at $\{z_1, \ldots, z_N\}$ with complex strengths $\{\strength_1, \ldots, \strength_N\}$, subject to a uniform flow with velocity $\Ustream = (U_{\infty}, V_{\infty})$ or in complex notation $\Wstream = U_{\infty} - i V_{\infty}$. We should stress the classical conjugation in the definition of the equivalent complex velocity for consistency with the Cauchy-Riemann equations \cite{eldredgebook}. The strengths are represented as $\strength_\Jidx = \flux_\Jidx - i \circu_\Jidx$ where $\flux_\Jidx,\; \circu_\Jidx \in \real{}$ denote the volume flux and the circulation of the $\Jidx$th singularity, respectively. By linearity, the solution of the Poisson equation is described by a complex potential $F = \pot + i \stream$, which at a location $z\in \complex{}$ is given by
\begin{equation}
\label{eqn:complexpotential}
    F(z) = \sum_{\Jidx = 1}^N \frac{S_\Jidx}{2\pi} \log(z - z_\Jidx) + \Wstream z;
\end{equation}
the velocity field $\vel$ is simply the derivative with respect to $z$. One can show that the dynamics of the singularities are governed by the following set of ordinary differential equations: 
\begin{equation}
\label{eqn:biotsavart}
    \ddt{z_\Jidx}  = \overline{w_{-\Jidx}}(t) = \sum_{\Kidx = 1, \Kidx \neq \Jidx}^N \frac{\overline{S_\Kidx}}{2\pi} \frac{1}{\overline{z_\Jidx} - \overline{z_\Kidx}} + \overline{W_{\infty}}, \hspace{0.3cm}
    \ddt{S_\Jidx}   = 0, \hspace{0.3cm} \Jidx=1, \ldots, N,
\end{equation} 
where $w_{-\Jidx}$ in complex notation, or $\vel_{-\Jidx}$ in vector notation, is the Kirchhoff velocity of the $\Jidx$th singularity, obtained by evaluating the velocity field at singularity $\Jidx$ and omitting its own contribution.

For potential flows, the pressure can equivalently be computed by two means: from the inversion of the pressure Poisson equation or from the unsteady Bernoulli equation; see Supplementary Material \ref{secSM:pressure}. Each formulation provides a different perspective on the mechanisms at play in the pressure response. The pressure Poisson equation is formed by taking the divergence of the Euler equations: 
\begin{equation}
\label{eqn:pressurePoisson}
    \lap\left(p(\position, t) + \frac{1}{2}\rho ||\vel(\position, t)||^2 \right) = \rho \sum_{\Jidx = 1}^N \left[\flux_{\Jidx} \vel_{-\Jidx}  \cdot \nabla \delta(\position - \position_{\Jidx}) - \circu_\Jidx \BB{e}_z \cdot (\vel_{-\Jidx} \times \nabla \delta(\position - \position_\Jidx)) \right].
\end{equation}
Equivalently to~\eqref{eqn:pressurePoisson}, the pressure induced by point singularities can be obtained in closed form from the unsteady Bernoulli equation. For a fixed evaluation point $z^\prime \in \complex{}$, we get
\begin{equation}
\label{eqn:pressuresolution}
    p(z^\prime,t) - B(t) =  -\frac{1}{2}\rho  \left| \sum_{\Jidx = 1}^N \frac{S_\Jidx}{2\pi} \frac{1}{z^\prime - z_\Jidx} + W_\infty\right|^2
     + \rho \re{\sum_{\Jidx = 1}^N \frac{S_\Jidx}{2\pi} \frac{\overline{w_{-\Jidx}}(t)}{z^\prime - z_\Jidx}},
\end{equation}
where $B(t)$ is function of time only. Eqns.~\eqref{eqn:complexpotential} and~\eqref{eqn:pressuresolution} are two examples relating the state (\ie singularity positions and strengths) and observation (\ie velocity potential or pressure measurements) that serve as prototypes for the larger class of problems in eqn.~\eqref{eqn:epde} that motivate this work. From their elliptic nature, we expect long-range nonlinear interactions between the state and observation. Furthermore,~\eqref{eqn:pressuresolution} shows that, by the appearance of the Kirchhoff velocity, every vortex affects the manner in which every other vortex induces pressure. For these reasons, estimating the characteristics of the singularities from limited observations poses many challenges.

\section{Low-rank ensemble Kalman filter}
\label{sec:enkf}

Before we present the traditional ensemble Kalman filter and our low-rank extension, we introduce our conventions. Serif fonts refer to random variables, \eg $\BB{\mathsf{Q}}$ on $\real{n}$ or $\mathsf{Q}$ on $\real{}$. Lowercase roman fonts refer to realizations of random variables, \eg $\BB{q}$ on $\real{n}$ or $q$ on $\real{}$. $\pdf{\BB{\mathsf{Q}}}$ denotes the probability density function for the random variable $\BB{\mathsf{Q}}$, and $\BB{q} \sim \pdf{\BB{\mathsf{Q}}}$ means that $\BB{q}$ is a realization of $\BB{\mathsf{Q}}$.  $\mean{\BB{\mathsf{Q}}}$ and $\cov{\BB{\mathsf{Q}}}$ denote the mean and the covariance matrix of the random variable $\BB{\mathsf{Q}}$, respectively. $\cov{\BB{\mathsf{Q}}, \BB{\mathsf{R}}}$ denotes the cross-covariance matrix of the random variables $\BB{\mathsf{Q}}$ and $\BB{\mathsf{R}}$. Empirical quantities are differentiated by carets above each symbol. 

A generic nonlinear discrete state-space model can be described by a pair of a Markov processes for a latent state variable $(\State_k)_{k \geq 0} \in \real{n}$, and an observed stochastic process $(\Meas_k)_{k \geq 1} \in \real{d}$ that is conditionally independent given the states at all times. The evolution of the state $\State_k$ is given by an initial distribution $\pdf{\State_0}$ and a \textit{dynamical system} to propagate the state from one time step to the next, \ie
\begin{equation}
\label{eqn:dynamic}
    \State_{k} = \dyn_k(\State_{k-1}) + \Noisedyn_k,
\end{equation}
where $\dyn_k\colon \real{n} \to \real{n}$ is a deterministic function that represents the mean of the discrete-time dynamics and $\Noisedyn_k$ is an independent \textit{process noise}. In our case, the state $\State_k$ is composed of the positions and strengths of a collection of point singularities, while $\dyn_k$ corresponds to the time-marching of their positions and strengths from discretizing~\eqref{eqn:biotsavart}. In this work, the observations  are given as a nonlinear mapping of the state $\State_k$ with additive noise of the following form:
\begin{equation}
\label{eqn:obs}
    \Meas_k = \obs_k(\State_k) + \Noiseobs_k,
\end{equation}
where $\obs_k :\real{n} \to \real{d}$ is the \textit{observation operator} and $\Noiseobs_k$ is an \textit{observation noise}. Eq.~\eqref{eqn:obs} is called \textit{the observation model}. In our case, the observation $\Meas_k$ is a vector of velocity potential or pressure evaluations at $d$ discrete locations.

\subsection{Ensemble Kalman filter}
For the remainder of this section, we focus on a single analysis step of the \enkf{}. Our treatment of the analysis step is built on the idea that there is an underlying transformation $\tmap_k$, called the \textit{prior-to-posterior transformation} or \textit{analysis map}, that directly maps samples from the forecast (\ie prior) density $\pdf{\State_k \given \Meas_{1:k-1} = \meas_{1:k-1}}$ to the filtering (\ie posterior) density $\pdf{\State_k \given \Meas_{1:k} = \meas_{1:k}}$~\cite{marzouk2016sampling, spantini2019coupling}. More precisely, $\tmap_k: \mathbb{R}^{n+d} \to \mathbb{R}^n$ maps the \textit{joint forecast distribution} of observations and states, $\pdf{\Meas_k, \State_k \given \Meas_{1:k-1} = \meas_{1:k-1}}$, to the filtering distribution $\pdf{\State_k \given \Meas_{1:k} = \meas_{1:k}}$. Further details on the filtering problem and the analysis map can be found in Supplementary Material \ref{secSM:filtering}.

Linear transform filters \cite{luenberger1997optimization} apply the following linear map to random variables $(\Meas_k, \State_k)$ representing the joint forecast distribution $\pdf{\Meas_k, \State_k \given \Meas_{1:k-1} = \meas_{1:k-1}}$, 
\begin{equation}
\label{eqn:enkfmap}
    \tmap_k(\Meas_k, \State_k) = \State_k - \cov{\State_k, \Meas_k} \cov{\Meas_k}^{-1}(\Meas_k - \meas_k^\star),
\end{equation}
where $\meas_k^\star$ is the observation to be assimilated, $\cov{\State_k, \Meas_k} \in \real{n \times d}$ is the cross-covariance matrix of the state and observation, while $\cov{\Meas_k}^{-1} \in \real{d \times d}$ is the precision matrix (inverse of the covariance matrix) of the observation's marginal distribution. The linear operator $\K_k =  \cov{\State_k, \Meas_k} \cov{\Meas_k}^{-1} \in \real{n \times d}$ is called the Kalman gain and maps observation discrepancies $(\meas_k - \meas_k^\star)$ to the state correction. 
The mean and covariance of $\tmap_k$ coincide with the standard Kalman update in the case of a Gaussian forecast distribution and linear--Gaussian observation model \cite[Corollary 1]{luenberger1997optimization}. In other words, the transformation yields the exact Bayesian update in the case of jointly Gaussian $(\Meas_k, \State_k)$. More generally, however, the mean of $\tmap_k$ is the linear least-squares estimate of the state at time $k$.
The \textit{stochastic ensemble Kalman filter (sEnKF)} introduced by Evensen~\cite{evensen1994sequential}  estimates the transformation~\eqref{eqn:enkfmap} by replacing the covariances with empirical covariances that are computed from the joint samples of the observations and states $\{\meas_k^i, \state_k^i\}$ $\sim \pdf{\Meas_k, \State_k \given \meas_{1:k-1}}$, \ie 
\begin{equation}
\label{eqn:enkfmap2}
    \widehat{\tmap}_k(\meas_k, \state_k) = \state_k - \widehat{\Sigma}_{\State_k, \Meas_k} \widehat{\Sigma}_{\Meas_k}^{-1}(\meas_k - \meas_k^\star).
\end{equation}

\subsection{\label{subsec:lowrank}Assimilation in low-dimensional subspaces}
For elliptic inverse problems, the conditional structure of the joint density of the observation and the state $\pdf{\Meas_k, \State_k \given \meas_{1:k-1}}$ is not localized, \ie there is no rapid decay of the correlations as a function of the distance between variables. Hence, distance based localization in this problem results in  biased estimators for the Kalman gain. Instead, we will exploit another kind of low-dimensional structure in the joint distribution. Our treatment draws inspiration from the fast multipole method (FMM)~\cite{greengard1987fast}, which uses a hierarchical clustering of ``source'' and
``target'' elements to accelerate the calculation of the potential field from a large set of point singularities~\cite{ethridge2001new}. These clusterings are typically based only on spatial distance, but here we will use information from the prior distribution and observation model (\ie observation operator and observation noise) to infer these clusters automatically in a way that works independently of the spatial distribution of the singularities and observation locations. To regularize the \enkf{}, we identify important directions in the state and observation spaces, perform the assimilation in these low-dimensional subspaces, and finally, lift the result to the original state space. First, we present the treatment of low-rank structure in the case of a linear--Gaussian observation model in \ref{subsec:lowrank}\eqref{subsubsec:linearlowrank}. The main result of this derivation is a factorization of the Kalman gain that exploits the existence of this low-dimensional subspace. Then, we extend this decomposition to the nonlinear--Gaussian setting in \ref{subsec:lowrank}\eqref{subsubsec:nonlinearlowrank}. An algorithm that summarizes the overall low-rank assimilation procedure is provided in  Supplementary Material \ref{secSM:algolowrank}.\\

\noindent\textbf{Remark: For convenience, we drop the time-dependent subscripts from the variables in the rest of this paper, since the analysis step does not involve time propagation.}

\subsubsection{Low-rank assimilation for the linear--Gaussian case}
\label{subsubsec:linearlowrank}
In this section, we consider the inference problem for the linear--Gaussian observation model:   
\begin{equation}
\label{eqn:linobs}
\Meas = \Obs \State + \Noiseobs,
\end{equation}
where the state is given by $\State \sim \N(\mean{\State}, \cov{\State})$, the observational error is given by $\Noiseobs \sim \N(\zero{}, \cov{\Noiseobs})$,  and $\Noiseobs$ is independent of $\State$. The matrix  $\Obs \in \real{d \times n}$ is called the observation matrix. In order to identify the important assimilation directions, we first define the whitened variables $\tilde{\State} = \cov{\State}^{-1/2} \left(\State - \mean{\State} \right) \in \real{n}$, $\tilde{\Noiseobs} = \cov{\Noiseobs}^{-1/2}\Noiseobs \in \real{d}$. We use $\BB{B}^{1/2}, \BB{B}^{-1/2}$ to denote a square root of the matrix $\BB{B}$ and its inverse, respectively. These whitened variables satisfy $\tilde{\Noiseobs} \sim \N(\zero{d}, \id{d}),\; \tilde{\State} \sim \N(\zero{n}, \id{n})$. Applying the same whitening to the observation variable, the observation model becomes 
\begin{equation}
\label{eqn:whitenedlik}
\tilde{\Meas} = \cov{\Noiseobs}^{-1/2} \left(\Meas - \mean{\Meas}\right) = \tilde{\Obs} \tilde{\State} + \tilde{\Noiseobs},
\end{equation}
where $\mean{\Meas} = \Obs \mean{\State}$ and $\tilde{\Obs} = \cov{\Noiseobs}^{-1/2} \Obs \cov{\State}^{1/2} \in \real{d \times n}$ is the whitened observation matrix. In this derivation we assume that $d \leq n$. We can now identify the important assimilation directions with a singular value decomposition (SVD) of the whitened observation matrix $\tilde{\Obs} = \BB{U} \BB{\Lambda} \BB{V}^\top$,  where $\BB{U} \in \real{d \times d}$ and $\BB{V} \in \real{n \times d}$ are the left and right singular vectors, and $\BB{\Lambda} \in \real{d \times d}$ is the diagonal matrix of singular values. From this decomposition, we can rotate and project the whitened state on the subspace spanned by the columns of $\BB{V}^\top$ --- $\breve{\State} = \BB{V}^\top \tilde{\State} \in \real{d}$ --- and similarly rotate the whitened observational error and observation on the subspace spanned by the columns of $\BB{U}^\top$ --- $\breve{\Noiseobs} = \BB{U}^\top \tilde{\Noiseobs} \in \real{d}, \; \breve{\Meas} = \BB{U}^\top \tilde{\Meas} \in \real{d}$. The SVD of the whitened observation matrix simultaneously identifies a pair of orthogonal bases for the state and observation spaces and an ordering for these directions. In Supplementary Material \ref{secSM:linearlowrank}, we show that the Kalman gain in the rotated space (where $\breve{\State}$ and $\breve{\Meas}$ live) is given by: 
\begin{equation}
\label{eqn:rotatedkalmangain}
    \breve{\K} = \BB{\Lambda}(\BB{\Lambda}^2 + \id{d})^{-1}.
\end{equation}
Thus, the linear analysis map in the rotated space $\breve{\tmap}$ is given by:
\begin{equation}
\label{eqn:rotatedlinearmap}
    \breve{\tmap}(\breve{\meas}, \breve{\state}) = \breve{\state} - \breve{\K}(\breve{\meas} - \breve{\meas}^{\star}) =  \breve{\state} - \BB{\Lambda}(\BB{\Lambda}^2 + \id{d})^{-1}(\breve{\meas} - \breve{\meas}^{\star}),
\end{equation}
where $\breve{\meas}^{\star} = \BB{U}^\top \cov{\Noiseobs}^{-1/2} \meas^{\star}$ denotes the realization of the assimilated data in the rotated observation space. In these rotated coordinates, the analysis map is local, \ie each component of the rotated observation only updates one associated component of the rotated state. Using~\eqref{eqn:rotatedkalmangain} and~\eqref{eqn:rotatedlinearmap}, we show in \ref{secSM:linearlowrank} that the Kalman gain $\K \in \real{n \times d}$ in the original space factorizes as
\begin{equation}
\label{eqn:kalmangainlowrank}
    \K = \cov{\State}^{1/2}\BB{V}\BB{\Lambda}(\BB{\Lambda}^2 + \id{d})^{-1}\BB{U}^\top \cov{\Noiseobs}^{-1/2}.  
\end{equation}
This factorization of $\K$ and its application in the analysis map of the Kalman filter~\eqref{eqn:enkfmap} provide a concise summary of the inference process in the low-rank informative subspaces: the innovation term $(\meas - \meas^\star)$ is whitened and rotated by applying $\BB{U}^\top \cov{\Noiseobs}^{-1/2}$, assimilated in this new set of coordinates using~\eqref{eqn:rotatedkalmangain}, and finally lifted to the original state space by applying $\cov{\State}^{1/2}\BB{V}$. In the whitened space (where $\tilde{\State}$ and $\tilde{\Meas}$ live), it is easy to see that $\BB{V}\BB{\Lambda}(\BB{\Lambda}^2 + \id{d})^{-1}\BB{U}^\top$ constitutes the singular value decomposition of the Kalman gain. From the decay of the singular values $\BB{\Lambda}$, we can instead use a truncated singular value decomposition of $\tilde{\Obs} \approx \BB{U}_r \BB{\Lambda}_r \BB{V}_r^\top$, where $r \leq d$, $\BB{U}_r \in \real{r\times r}, \BB{V}_r \in \real{n \times r}$ are the first $r$ left and right singular vectors, and $\BB{\Lambda}_r \in \real{r \times r}$. A rank$-r$ approximation of the Kalman gain is then given by $\K_r = \cov{\State}^{1/2}\BB{V}_r\BB{\Lambda}_r(\BB{\Lambda}_r^2 + \id{r})^{-1}\BB{U}_r^\top \cov{\Noiseobs}^{-1/2}$. We should emphasize that, without the truncation of the SVD, eq.~\eqref{eqn:kalmangainlowrank} is an exact factorization of the Kalman gain. With the truncated SVD, the state and observation variables are no longer just rotated but also projected to a subspace of dimension $r \leq \min(d,n)$.

\subsubsection{Low-rank assimilation for the nonlinear--Gaussian case}
\label{subsubsec:nonlinearlowrank}

In this section, we consider the nonlinear observation model of eq.~\eqref{eqn:obs}, recalled for reference:  
\begin{equation}
\label{eqn:nonlinearobs}
\Meas = \obs(\State) + \Noiseobs,
\end{equation}
where the state is $\State \sim \N(\mean{\State}, \cov{\State})$, the observational error is given by $\Noiseobs \sim \N(\zero{}, \cov{\Noiseobs})$, and $\Noiseobs$ is independent of $\State$. To handle a nonlinear observation operator with the linear prior-to-posterior transformation \eqref{eqn:enkfmap}, one simple approximation is to use the Jacobian of the observation operator about the prior mean as the observation matrix~\cite{asch2016data}. Unfortunately, this treatment is only viable for functions which are well approximated by a linear function over the bulk of the prior distribution. This section generalizes the treatment of the previous section for a nonlinear--Gaussian observation model; see Supplementary Material \ref{secSM:nonlinearlowrank} for further details.

In the nonlinear and non-Gaussian setting, Cui et al.~\cite{cui2021data} showed that the most important assimilation directions in the whitened state space can be identified by the eigenvectors of the \textit{state space Gramian}, which in the case of Gaussian $\Noiseobs$ reduces to:
\begin{equation}
\label{eqn:cx}
    \Cx = \int \left(\cov{\Noiseobs}^{-1/2}\nabla \obs(\state) \cov{\State}^{1/2}\right)^\top \left(\cov{\Noiseobs}^{-1/2}\nabla \obs(\state) \cov{\State}^{1/2}\right)  \mathsf{d}\pdfprior(\state)\in \real{n\times n},
\end{equation}
where the expectation is taken over the prior distribution. As expected,  eq.~\eqref{eqn:cx} reverts to $\tilde{\Obs}^\top \tilde{\Obs}$ in the linear--Gaussian case. $\Cx$ is \textit{positive semi-definite (\psd{})} and its eigendecomposition can be written as
$\Cx = \BB{V} \BB{\Lambda}_{\State}^2\BB{V}^\top$, where $\BB{V} \in \real{n\times n}$ is an orthogonal basis for the whitened state space with associated eigenvalues $\BB{\Lambda}^2_{\State} \in \real{n\times n}$.

 Inspired by the treatment in the state space, we propose to use the eigenvectors of the \textit{observation space Gramian} $\Cy$ to select the important assimilation directions in the whitened observation space:
\begin{equation}
\label{eqn:cy}
    \Cy  = \int \left(\cov{\Noiseobs}^{-1/2}\nabla \obs(\state) \cov{\State}^{1/2}\right) \left(\cov{\Noiseobs}^{-1/2}\nabla \obs(\state) \cov{\State}^{1/2}\right)^\top  \mathsf{d}\pdfprior(\state) \in \real{d \times d}.
\end{equation}
The matrix $\Cy$ is also \psd{} and has the eigendecomposition
$\Cy = \BB{U} \BB{\Lambda}^2_{\Meas}\BB{U}^\top$, where $\BB{U}\in \real{d \times d}$ is an orthonormal basis for the whitened observation space with associated eigenvalues $\BB{\Lambda}^2_{\Meas} \in \real{d \times d}$. For convenience, we assume that the eigenvectors of $\Cx$ and $\Cy$ are ordered by decreasing eigenvalues.  For a nonlinear observation model, the eigenvalues of the state and observation Gramians can be different. In practice, we use Monte-Carlo approximations of $\Cx$ and $\Cy$ that are estimated using the prior samples to identify the important subspaces. Depending on the inference problem, the Jacobian $\nabla \obs \in \real{d \times n}$ of the observation operator $\obs$ with respect to the state components can be computed either analytically, with automatic differentiation, with complex step differentiation, or, in the worst case, from finite differences. To perform the low-rank assimilation, we only retain the first $\rx \leq n$ eigenmodes for the state space, and $\ry \leq d$ eigenmodes for the observation space. The ranks $\rx$ and $\ry$ can be tuned independently based on the decay of $\BB{\Lambda}_{\State}^2$ and $\BB{\Lambda}_{\Meas}^2$; typically we recommend setting these ranks to achieve a threshold $\alpha~\in~[0, 1]$ for the cumulative normalized energy of the eigenvalue spectra, \eg in the state space, we set $\rx = \min{\{r_{\alpha} \in \llbracket 1, n \rrbracket \given   E_{\State, r_\alpha} \geq \alpha\}}$, where $E_{\State, r_\alpha} = \sum_{i=1}^{r_\alpha}\lambda^2_{\State, i}/\sum_{i=1}^n\lambda^2_{\State, i}$. The new projected state and projected observation variables are defined as $\breve{\State}  = \BB{V}^\top_{\rx} \tilde{\State} = \BB{V}^\top_{\rx} \cov{\State}^{-1/2}(\State - \mean{\State}) \in \real{\rx}$ and $\breve{\Meas}  = \BB{U}^\top_{\ry} \tilde{\Meas} = \BB{U}^\top_{\ry} \cov{\Meas}^{-1/2}(\Meas - \mean{\Meas}) \in \real{\ry}$, where $\BB{V}_{\rx}, \BB{U}_{\ry}$ denote the first $\rx, \ry$ columns of $\BB{V}, \BB{U}$, respectively.

We now outline how to obtain the factorization of the Kalman gain in the nonlinear setting. In the linear--Gaussian case, the Gramians in~\eqref{eqn:cx} and~\eqref{eqn:cy} revert to $\Cx = \tilde{\Obs}^\top \tilde{\Obs}$ and $\Cy = \tilde{\Obs} \tilde{\Obs}^\top$, respectively. From the definition of the SVD of the whitened observation matrix $\tilde{\Obs}$, we have $\tilde{\Obs} \BB{v}_i = \lambda_i \BB{u}_i$ for some left singular vector $\BB{u}_i$, singular value $\lambda_i$ and right singular vector $\BB{v}_i$. This results in the eigendecompositions $\Cx u_i = \lambda_i^2 \BB{u}_i$ and $\Cy \BB{v}_i = \lambda_i^2 v_i$. Therefore, if the triplet $\{ \BB{u}_i, \lambda_i, \BB{v}_i \}$ is obtained from the eigendecomposition of $\Cx$ and $\Cy$, the sign of the singular vectors is lost, \ie we only have $\tilde{\Obs} \BB{v}_i = \pm \lambda_i \BB{u}_i$. For a nonlinear observation model, the factorization of the Kalman gain given in eq.~\eqref{eqn:kalmangainlowrank} is no longer applicable and needs to be generalized. In the rotated and whitened space, the analysis map is given by: 
\begin{equation}
    \breve{\tmap}(\breve{\meas}, \breve{\state})  =
    \breve{\state} - \Kbreve(\breve{\meas} - \breve{\meas}^\star),
\end{equation}
where $\Kbreve  = \cov{\breve{\State} \breve{\Meas}} \cov{\breve{\Meas}}^{-1} \in \real{\rx \times \ry}$ is the Kalman gain in the informative space with $\cov{\breve{\State} \breve{\Meas}} \in \real{\rx \times \ry}$ and $\cov{\breve{\Meas}}^{-1} \in \real{\ry \times \ry}$. In the original space, we get
\begin{equation}
\label{eqn:kalmangain_nonlinear}
    \tmap(\meas, \state) = \state - \cov{\State}^{1/2}\V \Kbreve \U^\top\cov{\Noiseobs}^{-1/2}(\meas - \meas^\star).
\end{equation}
This new factorization of the Kalman gain nicely generalizes eq.~\eqref{eqn:kalmangainlowrank} to an inference problem with a nonlinear observation model. We emphasize that the sign issue presented above is obviated by absorbing it into  the definition of $\Kbreve$. Let us remark that the definition of $\Kbreve$ reduces to the diagonal matrix $\BB{\Lambda}(\BB{\Lambda}^2 + \id{})^{-1}$ in the linear--Gaussian case, where the matrices $\U, \V$ are obtained from the SVD of $\tObs$. Indeed, eq.~\eqref{eqn:kalmangain_nonlinear} constitutes \textit{a change of coordinates for the Kalman gain} between the original space and the informative space. 
We denote the algorithm applying the linear update in~\eqref{eqn:kalmangain_nonlinear} in each analysis step as the low-rank \enkf{} (\lrenkf{}). A pseudo-code for the proposed low-rank \enkf{} is presented in \ref{secSM:algolowrank}.

In comparison, the \senkf{} estimates the Kalman gain in the original space, a linear operator of dimensions $n \times d$, where $n$ and $d$ are the dimensions of the space and observation spaces. Leveraging the dimension reduction offered by the state and observation Gramians, the \lrenkf{} only has to estimate the Kalman gain in the informative subspaces of $\State$ and $\Meas$, whose dimension is only $\rx \times \ry$, where $\rx$ and $\ry$ can be significantly smaller than $n$ and $d$, respectively. To appreciate the benefit of this dimension reduction in the inference problem, it is useful to recall the classical bias-variance trade-off in any statistical learning problem~\cite{friedman2001elements}: Given an ensemble size $M$, the error in the estimation of the Kalman gain $\K$ can be decomposed into a variance term and a bias  term. For a limited ensemble size (compared to $n$ and $d$), the estimated Kalman gain from the vanilla \senkf{} will have a large variance, potentially leading to a  filter with diverging state estimation error, as is the case for small ensemble size in the example problems. We show later in this paper that with the proposed dimension reduction, the Kalman gain in the informative subspaces $\Kbreve$ contains a much smaller number of entries than $\K$. This results in smaller variance. The bias in computing $\Kbreve$ can be controlled by setting the ranks of the state and observation projections. Due to the rapid decay of the spectrum of the state and observation Gramians, increasing the rank of the projected subspace beyond a certain ratio of the cumulative energy will only marginally reduce the bias, but greatly increase the variance. In other words, we choose the ranks $\rx, \ry$ to capture the column/row space of the Kalman gain and make the bias  small for any finite-sample-estimator of $\K$. Finally, we should stress that without dimension reduction (\ie $\rx = n$ and $\ry =d$), the \senkf{} and the \lrenkf{} are equivalent, up to the rotations of the variables.
Indeed, the \senkf{} estimates the Kalman gain $\K \in \real{n\times d}$ in the original coordinates, while the \lrenkf{} estimates $\Kbreve \in \real{n \times d}$ in the informative coordinates and lifts it to the original coordinates according to \eqref{eqn:kalmangain_nonlinear}.

\subsection{\label{subsec:daviskahan}Estimation of the leading directions from samples}
In practice, we only have access to limited samples $\{\state^1, \ldots, \state^M\}$ from the prior distribution to estimate the state and observation Gramians $\Cx, \Cy$ and their associated eigenvectors. To understand how well we can approximate these leading eigenvectors from the empirical Gramians $\sCx, \sCy$, we recall the following corollary of the Davis--Kahan theorem ~\cite{davis1970rotation, yu2015useful}:

\begin{theorem}[Corollary of the Davis--Kahan theorem 
\cite{davis1970rotation, yu2015useful}\footnote{A slightly more general version is presented in \cite{yu2015useful}.} \label{thm:davis}]
Let $\BB{\Gamma}$ and $\widehat{\BB{\Gamma}}$ be $q \times q$ positive semi-definite matrices with eigendecompositions $\BB{\Gamma} \BB{w}_i = \lambda_i^2 \BB{w}_i$ and $\widehat{\BB{\Gamma}} \hat{\BB{w}}_i = \hat{\lambda}_i^2 \hat{\BB{w}}_i$, respectively, with eigenvalues sorted in descending order: $\lambda_1^2 \geq \lambda_2^2 \geq \ldots \geq \lambda_m^2 \geq 0$, $
\hat{\lambda}_1^2 \geq \hat{\lambda}_2^2 \geq \ldots \geq \hat{\lambda}_q^2 \geq 0$. For $1 \leq r < q$ with $\lambda_r^2 - \lambda_{r+1}^2 > 0$, we define $\BB{W}_r = \left(\BB{w}_1, \BB{w}_{2}, \ldots, \BB{w}_r\right) \in \real{q \times r}$ and $\widehat{\BB{W}}_r = \left(\hat{\BB{w}}_1, \hat{\BB{w}}_{2}, \ldots, \hat{\BB{w}}_r\right) \in \real{q \times r}$. Then, the distance between the subspaces spanned by the columns of $\BB{W}_r$ and $\widehat{\BB{W}}_r$, denoted $d(\BB{W}_r, \hat{\BB{W}}_r)$, satisfies
\begin{equation}
    d(\BB{W}_r, \widehat{\BB{W}}_r) \coloneqq \left|\left|\BB{W}_r \BB{W}_r^\top \left(\id{} - \widehat{\BB{W}}_r \widehat{\BB{W}}_r^\top\right)\right|\right|_F \leq \frac{\|\BB{\Gamma} - \widehat{\BB{\Gamma}} \|_{F} }{\lambda^2_{r} - \lambda^2_{r+1}} .
\end{equation}
\end{theorem}

Let us now apply the Davis--Kahan theorem to the pairs $(\BB{\Gamma}, \widehat{\BB{\Gamma}}) = (\Cx, \sCx)$, and $(\Cy, \sCy)$.  Two important conclusions can be drawn from this result. First, it follows from Theorem \ref{thm:davis} that the error in the subspace spanned by the first $r$ eigenvectors of $\widehat{\BB{\Gamma}}$ converges to zero at the same rate that  $\widehat{\BB{\Gamma}}$ converges to $\BB{\Gamma}$ in the Frobenius norm, as $M \rightarrow \infty$. See Tropp~\cite{tropp2015introduction} for relevant results on the convergence rates of random matrices. Second, the error in the estimated $r$-dimensional leading eigenspace depends on the difference between consecutive eigenvalues  $\lambda_r^2 -  \lambda_{r+1}^2$, 
which is called the spectral gap. Specifically, the inverse of the spectral gap is a useful indicator of the estimation error in the informative subspaces. Intuitively, two eigenvectors with close eigenvalues are difficult to distinguish. For the fluid dynamics problems considered in this paper (see examples in Section~\ref{sec:examples}), the spectra of the state and observation Gramians decay rapidly, and hence the leading eigenvalues are well separated. This suggests that we can estimate the leading eigendirections of $\Cx, \Cy$ with relatively few samples $M$, and that the row and column spaces of the Kalman gain can be accurately captured using only a small number of such eigendirections.

\section{Examples \label{sec:examples}}

\subsection{The leading directions of the state and observation Gramians and the multipole expansion}
\label{subsec:CxCymultipole}

\begin{figure}
    \centering
    \includegraphics[width = 0.6\linewidth]{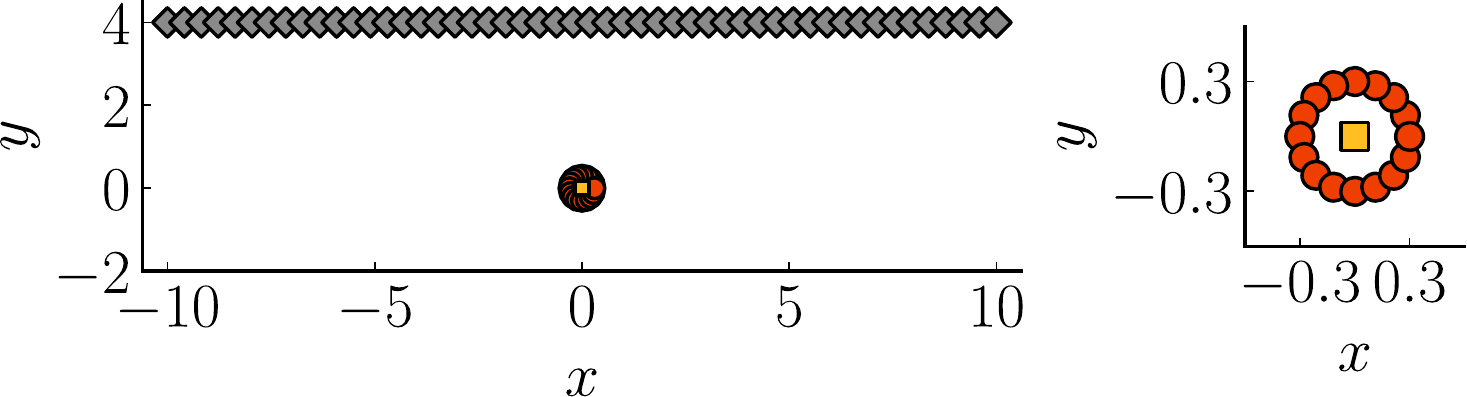}
    \caption{Schematic of the setup. Orange dots depict the location of the point sources. The yellow square (located at $(0, 0)$) depicts the location of the centroid of the point sources. Grey diamonds depict the location of evaluation points of the velocity potential $\pot$.}
    \label{fig:setup_example1}
\end{figure}

In this example, we develop intuition for the ideas presented in this paper by connecting the leading eigenvectors of the state and observation space Gramians discussed in eqns.~\eqref{eqn:cx} and~\eqref{eqn:cy} to the multipole expansion. We consider a set of $N$ point sources located at $\{z_1, \ldots, z_N\}$ with strengths $\{\flux_1, \ldots, \flux_N\}$, and a set of $d$ evaluation points located at $\{\zp_1, \ldots, \zp_d\}$.  We assume that the strengths are perfectly known, but we seek to estimate the positions of the singularities from observations of the velocity potential $\phi$ at the evaluation points corrupted by additive Gaussian noise $\noiseobs \sim \N(\zero{d}, \cov{\noiseobs})$. To ensure a convergent multipole expansion, we place all the evaluation points outside a circle of radius $R$ where $R = \max_{\Jidx}|z_{\Jidx}|$. We use the following notations: $\state = [z_1, \ldots, z_N]^\top \in \complex{N}$, $\BB{S} = [\flux_1, \ldots, \flux_N]^\top \in \real{N}$ and $\BB{\zp} = [\zp_1, \ldots, \zp_d]^\top \in \complex{d}$. The superscript $\top$ denotes the transpose of a complex vector/matrix without conjugation, while the superscript $\mathsf{H}$ denotes the transpose with conjugation of a complex vector/matrix.

The $i$th component of the observation vector $\meas \in \real{d}$ is given by $ y_i = \pot(\zp_i; \state) + \epsilon_i, i = 1, \ldots, d$, where the notation $\pot(\zp_i; \state)$ is used to highlight the dependence on the positions of the point sources. We seek to estimate the position vector $\state$ from the noisy observations $\meas$. The observation model reads in vector form as $\meas = \obs(\state) + \noiseobs$, where $\obs \colon \complex{N} \to \real{d}, \; \state \mapsto [\pot(\zp_1; \state), \ldots, \pot(\zp_d; \state)]^\top$. From section~\ref{subsec:lowrank}, the informative subspaces are identified from the Jacobian of the observation operator $\obs$ with respect to the state variable $\state$, denoted $\nabla_{\state} \obs(\state)$. If $|z_\Jidx|/|\zp_k| <1$ for all $k, \Jidx$, then we get from~\eqref{eqn:complexpotential}

\begin{equation}
\label{eqn:multipole_factorization}
      \nabla_{\state} \obs = \frac{1}{4\pi} \begin{bmatrix}
    \frac{-\flux_1}{\zp_1 - z_1}&  \ldots & \frac{-\flux_N}{\zp_1 - z_N}\\
    \vdots & & \vdots \\
    \frac{-\flux_1}{\zp_d - z_1} & \ldots & \frac{-\flux_N}{\zp_d - z_N} 
    \end{bmatrix}
    \approx \underbrace{\begin{bmatrix}
    \frac{1}{\zp_1}&  \ldots & \frac{1}{{\zp_1}^p}\\
    \vdots & & \vdots \\
    \frac{1}{\zp_d}&  \ldots & \frac{1}{{\zp_d}^p}
    \end{bmatrix}}_{\BB{A}_{\BB{\zp}}}
  \underbrace{\begin{bmatrix}
  1 & \ldots & 1\\
  z_1 & \ldots & z_N\\
  \vdots & & \vdots \\
  z_1^p & \ldots & z_N^p
  \end{bmatrix}}_{{\BB{B}_{\state}}^\top} \diag{-\frac{\BB{S}}{4\pi}},
\end{equation}
where the approximation comes from truncating the multipole expansion at the $p$-th order. This factorization decouples the contribution of the position of the evaluation points ($\BB{A}_{\BB{\zp}}$), the position of the point sources ($\BB{B}_{\state}^\top$), and the volume fluxes of the point sources ($\diag{-\BB{S}}$). Unfortunately, it is not possible to compute analytically the Gramians $\Cx, \Cy$ for this inference problem, even with a Gaussian distribution for the $x$ and $y$ coordinates of the  point source locations. To highlight the connection between the eigenvectors of these matrices and the multipole expansion, we compare the singular vectors obtained from the SVD of $\nabla_{\state} \obs$ with the orthonormal bases obtained by orthonormalization of the matrices $\BB{A}_{\BB{\zp}}, \BB{B}_{\state}$ for a particular position vector $\state$. We recall that the left singular vectors $\BB{u}_i$ of $\nabla_{\state} \obs$ are also the eigenvectors of the Gram matrix $\nabla_{\state} \obs^{\mathsf{H}} \nabla_{\state} \obs$, and similarly, the right singular vectors $\BB{v}_i$ of $\nabla_{\state} \obs$ are also the eigenvectors of the Gram matrix $\nabla_{\state} \obs \nabla_{\state} \obs^{\mathsf{H}}$. To compute an orthonormal basis for the columns of $\BB{A}_{\BB{\zp}}$ and $\BB{B}_{\state}$, we extract the $\Q$ factor of the QR factorizations of these matrices, denoted by $\Q_{\BB{\zp}} \in \complex{d \times p}$ and $\Q_{\state} \in \complex{n \times p}$, respectively.

We place a set of $N = 16$ point sources distributed according to $z_\Jidx = \rho \exp{({2\pi i \Jidx}/{N})}$ for $\Jidx = 1, \ldots, N$, where $\rho = 0.3$. The volume flux of the point sources is set to $4 \pi$. The centroid of this set of point sources is at the origin. This is critical for a sensible comparison with the factorization of \eqref{eqn:multipole_factorization}. We use $d = 50$ evaluation points distributed along a horizontal line as $\zp_k = -10 + (k-1) \Delta s + 4 i$, for $k = 1, \ldots, d$ with interspace $\Delta s = 20/(d-1)$.  Fig.~\ref{fig:setup_example1} depicts the configuration of the different elements. We truncate the multipole expansion at the $p =30$th order to get a machine precision approximation of $\dobs$ in eq.~\eqref{eqn:multipole_factorization}.

\begin{figure}
\centering
    \includegraphics[width = 0.9\linewidth]{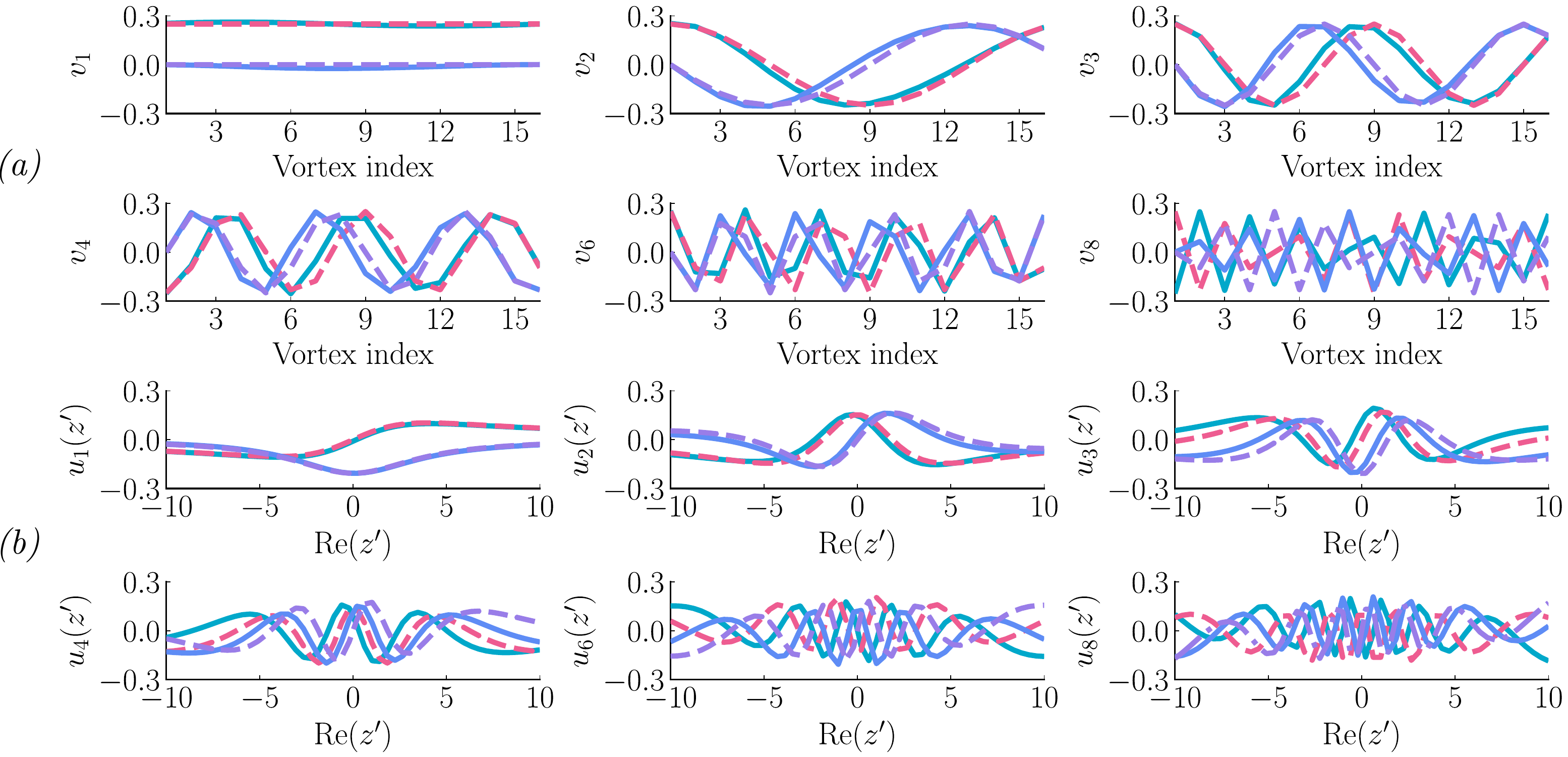}
\caption{Top panel (a): Comparison of the real and imaginary parts of the $1, 2,3, 4, 6$ and  $8$th state modes ($\BB{v}_i$s) obtained from the SVD of $\nabla_{\state} \obs$ (solid turquoise and solid blue lines), and from orthogonalization of the columns of $\Q_{\state}$ (dashed red and dashed purple lines). Lower panel (b): Comparison of the real and imaginary parts of the $1, 2,3, 4, 6$ and  $8$th observation modes ($\BB{u}_i$s) obtained from the SVD of $\nabla_{\state} \obs$ (solid turquoise and solid blue lines), and from the columns of $\Q_{\BB{\zp}}$ (dashed red and dashed purple lines). The vectors have unit norm.}
\label{fig:modes_sourcetarget}
\end{figure}

Fig.~\ref{fig:modes_sourcetarget} (a) compares the real and imaginary parts of the $1, 2,3, 4, 6$ and  $8$th modes obtained from the right singular vectors of $\nabla_{\state} \obs$, and from the corresponding columns of the unitary matrix $\Q_{\state}$. Fig.~\ref{fig:modes_sourcetarget} (b) compares the real and imaginary parts of the $1, 2,3, 4, 6$ and  $8$th modes obtained from the left singular vectors of $\nabla_{\state} \obs$, and from the corresponding columns of the unitary matrix $\Q_{\BB{\zp}}$. The first state mode $\BB{v}_1$ is $\pm \BB{1}/\sqrt{N}$, where $\BB{1}$ denote a vector of ones of length $N$. Therefore, the leading state correction corresponds to a rigid body translation of the point sources.  Three modes capture more than $99.999999\%$ of the cumulative energy of $\dobs^\mathsf{H} \dobs$.  Overall, the state and observation modes obtained from these two procedures show strong similarities, especially for the first three modes. However, the modes are not expected to coincide exactly. The differences in the vectors obtained from these two approaches may be attributed to $\Q_{\BB{\zp}}$ only relying on the positions of the evaluation points, and $\Q_{\state}$ only relying on the positions of the point sources. In contrast, the SVD simultaneously has access to the locations of the evaluation points and the point sources to identify the left and right singular vectors. As a result, the SVD constructs an \textit{optimized} ``multipole expansion'' of the Jacobian (see the Eckart--Young theorem \cite{eckart1936approximation}), specifically designed for the particular set of the evaluation points and the point sources. 

\subsection{Inference of the properties of point vortices from pressure observations along a wall\label{subsec:pointvortex}}

\begin{figure}
    \centering
    \includegraphics[width = 0.9\linewidth]{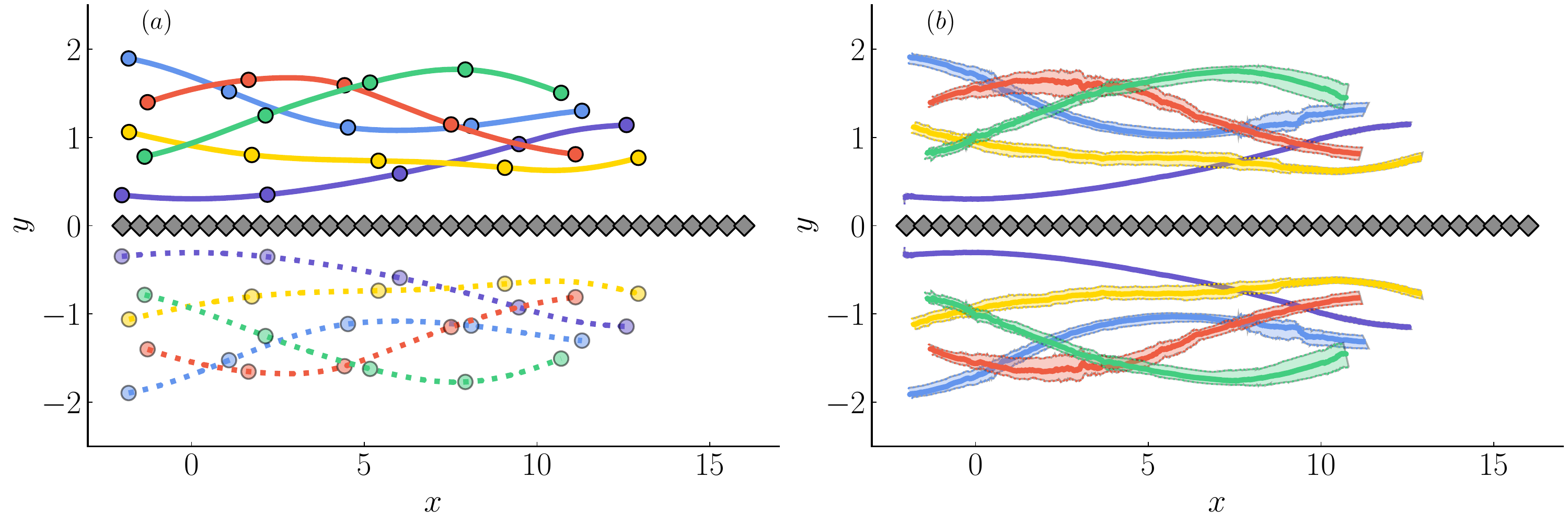}
    \caption{(a): Schematic of the setup. True trajectories of the point vortices (for one realization of the system) are represented by colored lines. The location of the point vortices, sampled every three convective times, is depicted by colored dots. Fainter dashed lines and fainter dots refer to the image point vortices. Location of the pressure sensors are depicted by grey diamonds. (b): Estimation of the trajectories of the vortices with the \lrenkf{} for $M =  40$ with the ranks $\rx$ and $\ry$ set to capture $99\%$ of the cumulative energy spectra. Solid lines depict the time history of the median posterior estimate for the position of the different point vortices. Fainted areas show the $5\%$ and $95\%$ quantiles of the posterior estimate for the position of the point vortices.}
    \label{fig:example2_trajectories}
\end{figure}

In this second example, we compare the \lrenkf{} with the \senkf{} for estimating the positions and strengths of a collection of point vortices advecting along a horizontal wall. We rely on pressure observations collected along the wall to estimate the state. These observations were generated from the same observation model used for inference, thereby making this a \textit{twin experiment} \cite{asch2016data}. We consider a set of $N = 5$ point vortices located at $\{z_1, \ldots, z_N\}$ with circulations $\{\circu_1, \ldots, \circu_N\}$, and a set of $d$ evaluation points located at $\{\zp_1, \ldots, \zp_d\}$, see Fig.~\ref{fig:example2_trajectories} (a). The state variable $\state$ contains the positions and circulations of the $N$ point vortices:  $\state = 
\left[x_{1},  y_{1},  \Gamma_{1},  \ldots  x_{N},  y_{N},  \Gamma_{N}
\right]^\top \in \real{3 N}$.
To avoid singular interactions between nearby vortices, we replace the singular Cauchy kernel $k(z) = 1/(2\pi z)$ used to compute the Kirchhoff velocities $w_{-\Jidx}$ in~\eqref{eqn:biotsavart} and~\eqref{eqn:pressuresolution} with the regularized algebraic blob kernel ${k}_{\epsilon}(z) = \overline{z}/(2\pi (|z|^2 + \epsilon^2))$, where $\epsilon$ is called the blob radius~\cite{eldredgebook}, set here to $5 \times 10^{-2}$. To enforce the no-flow-through condition along the $x$ axis, we use the method of images and augment our collection of vortices with another set of $N$ vortices at the conjugate positions $\{\overline{z_1}, \ldots, \overline{z_N}\}$ with opposite circulation $\{-\circu_1, \ldots, -\circu_N\}$. We emphasize that these mirrored vortices are only an artifice to enforce the no-flow-through in the forecast step, and play no role in the analysis step.  We add a freestream flow directed along increasing $x$ values with velocity $\Ustream = [1, 0]^\top$. 
The initial position of the $\Jidx$th point vortex is generated randomly with the form $z_{\Jidx} + \rho_r \exp(i \theta)$ where $\{z_1, \ldots, z_N\} = \{-2.0 + 0.3i, -1.9 + 1.9i, -1.8 + 1.1i, -1.3 + 1.4i, -1.4 + 0.8i \}$, $\rho_r$ is drawn from $\N(0.0, 0.1)$ and $\theta$ is drawn from the uniform distribution on $[0, \; \pi]$. The initial circulation of the point vortices is drawn from $\N(0.4, 0.1)$. The collection of point vortices is advanced using a forward Euler scheme with time step $\Delta t = 1\times 10^{-3}$. The observation vector $\meas$ consists of pressure observations at $d = 37$ locations linearly distributed with an interspace of $0.5$ along the segment $[-2, 16]$, see Fig.~\ref{fig:example2_trajectories} (a). We use a Gaussian observation noise with zero mean and covariance $1 \times 10^{-4}$. We obtain the Jacobian of the observation operator used in the state and observation Gramians $\Cx$ and $\Cy$ by analytical differentiation of eq.~\eqref{eqn:pressuresolution}.

\begin{figure}
    \centering
    \includegraphics[width = \linewidth]{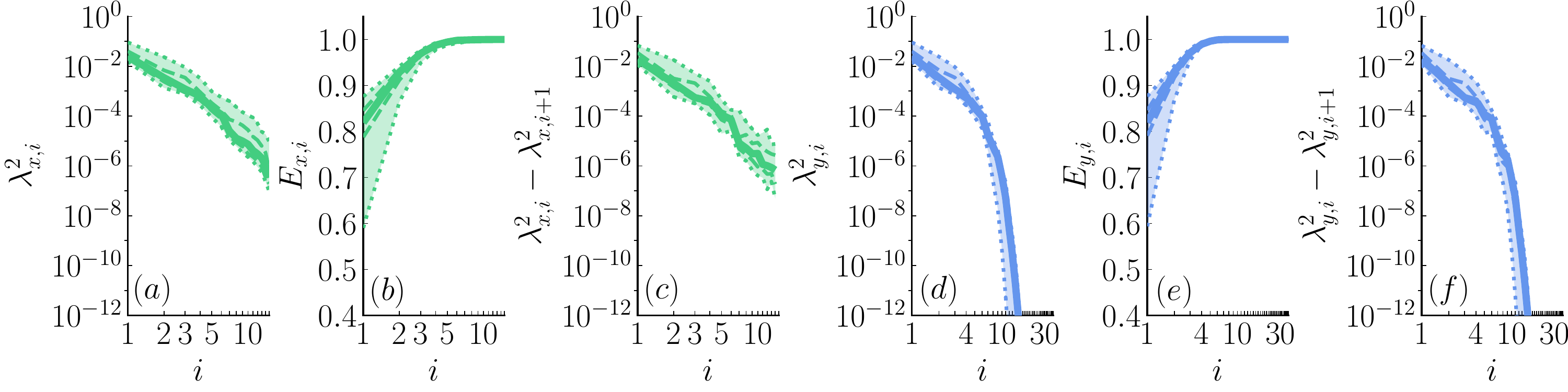}
    \caption{Median spectrum of the state Gramian $\Cx$ (left panels (a)-(c)) and observation Gramian $\Cy$ (right panels (d)-(f)) over the time interval $[0, 12]$. Statistics are obtained from a run of the \senkf{} with $M = 1000$. Panels [(a), (d)]: Median eigenvalues of $\Cx$ and $\Cy$. Panels [(b), (e)]: Median normalized cumulative energy $E_i = \sum_{j=1}^i\lambda_j^2/\sum_{j}\lambda_j^2$ of $\Cx$ and $\Cy$. Panels [(c), (f)]: Median spectral gap $\lambda_i^2 - \lambda_{i+1}^2$ of $\Cx$ and  $\Cy$. The abscissa axis is in log scale. Dashed lines depict the $25\%$ and $75\%$ quantiles. Dotted lines depict the $5\%$ and $95\%$ quantiles.}
    \label{fig:example2_energy_spectrum}
\end{figure}

We assess the performance of the two filters with a \textit{twin experiment}. We draw a random initial condition for the positions and strengths of the different point vortices. Then we simulate the dynamics of the point vortices over the time interval $[0, 12]$ (corresponding to $12000$ assimilation cycles with the choice of time step $\Delta t$). The true state at the time step $k$ is denoted by $\state^\star_k$. At every time step, we generate the noisy pressure observation at the sample locations based on the true state $\state^\star_k$. The realization of the true observation vector at the time step is denoted by $\meas_k^\star$. In the twin experiment, we seek to estimate the true state $\state^\star_k$ only from the knowledge of the noisy and indirect observations $\meas_k^\star$ and the prior.

\begin{figure}
    \centering
    \includegraphics[width = 0.8\linewidth]{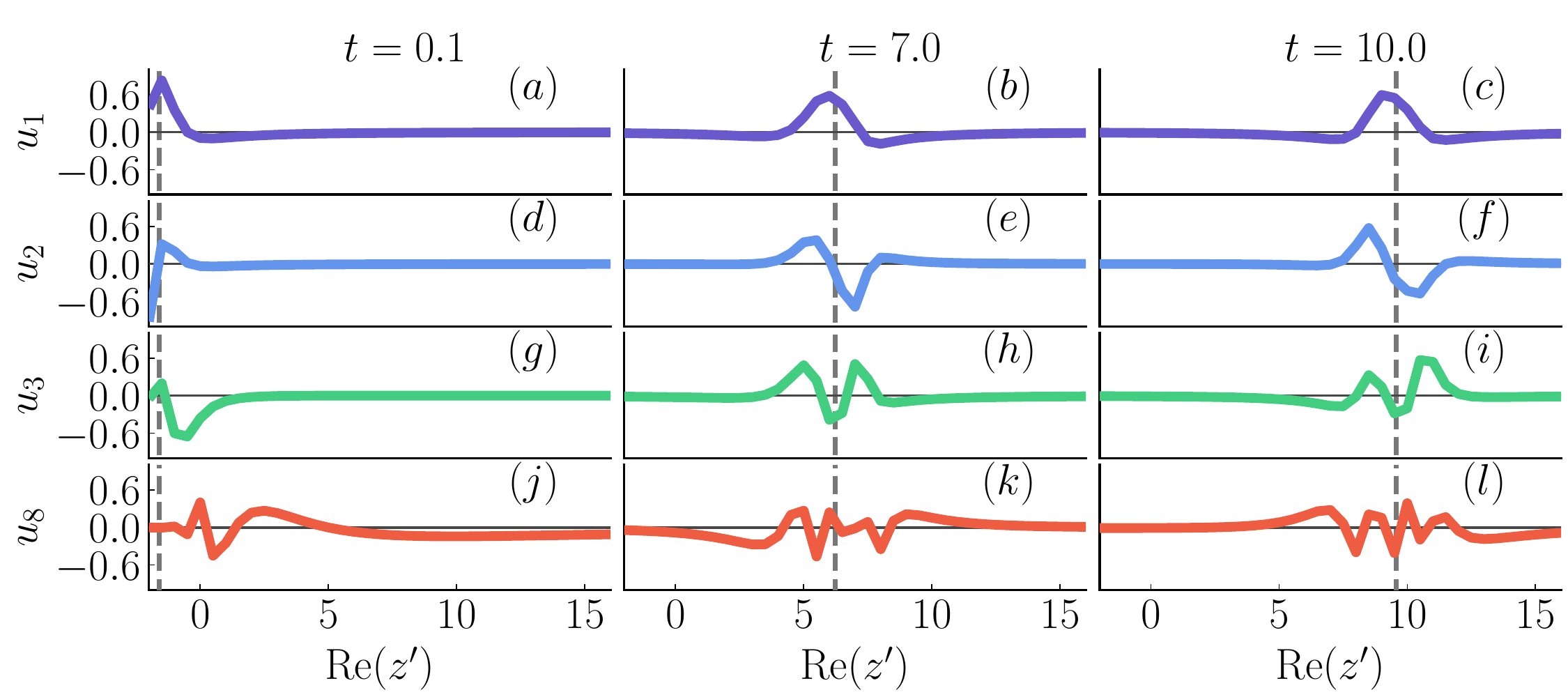}
    \caption{$1, 2, 3$ and $8$th eigenvectors of the observation Gramian $\Cy$, \ie observation modes $\BB{u}_i$s,  at $t = 0.1$ (left column), $t = 7.0$ (middle column), and $t = 10.0$ (right column). (a)-(c): first mode, (d)-(f): second mode, and (g)-(i): third mode, (j)-(l): eight mode. The dashed grey vertical line depicts the $x$ component of the position of the centroid of the vortices. The grey horizontal line corresponds to $0$ ordinate. Statistics are obtained from a run of the \senkf{} with $M = 1000$. The eigenvectors have unit norm.}
    \label{fig:example2_u_eigenvector}
\end{figure}

First, we assess the spectra and leading subspaces of the state and observation Gramians. Fig.~\ref{fig:example2_energy_spectrum} depicts the median eigenvalues of the state and observation Gramians over the time interval $[0, 12]$. We run the \senkf{} with ensemble size $M = 1000$ to generate the  samples used to compute empirical Gramians in each analysis step. We choose this larger ensemble size to reduce sampling errors. The median ranks $\rx$ to capture $80\%, 90\%, 95\%,$ and $99\%$ of the cumulative energy of $\Cx$ are $1, 2, 3,$ and $5$, respectively. Similarly, the median ranks $\ry$ to capture $80\%, 90\%,$ and $99\%$ of the cumulative energy of $\Cy$ are $1, 2,$ and $4$, respectively. The sharp decay of the spectra of $\Cx$ and $\Cy$ supports our hypothesis of low-rank structure existing in the prior-to-posterior update. A low-dimensional subspace of the pressure observations is only informative along a limited number of directions in the state space. Fig.~\ref{fig:example2_u_eigenvector} shows the $1, 2, 3$ and $8$th eigenvectors of the observation Gramian at the three times: $t = 0.1, 7.0,$ and $10.0$. These modes clearly illustrate how information is extracted from the different pressure discrepancies during the analysis step. The first observation mode corresponds to a spatially weighted average of the pressure discrepancies about the mean $x$ coordinate of the position of the vortices (depicted by the dashed vertical line on the panels of Fig.~\ref{fig:example2_u_eigenvector}). The second mode captures differences between the pressure observations collected on the left and right side of the centroid location. The higher modes follow the same pattern and act as refined stencils with growing support to extract higher order features from the pressure observations. The leading eigenvectors of the state Gramian are more difficult to interpret, due to the manner in which vortices are coupled in the determination of pressure.

\begin{figure}
    \centering
    \includegraphics[width = 0.8\linewidth]{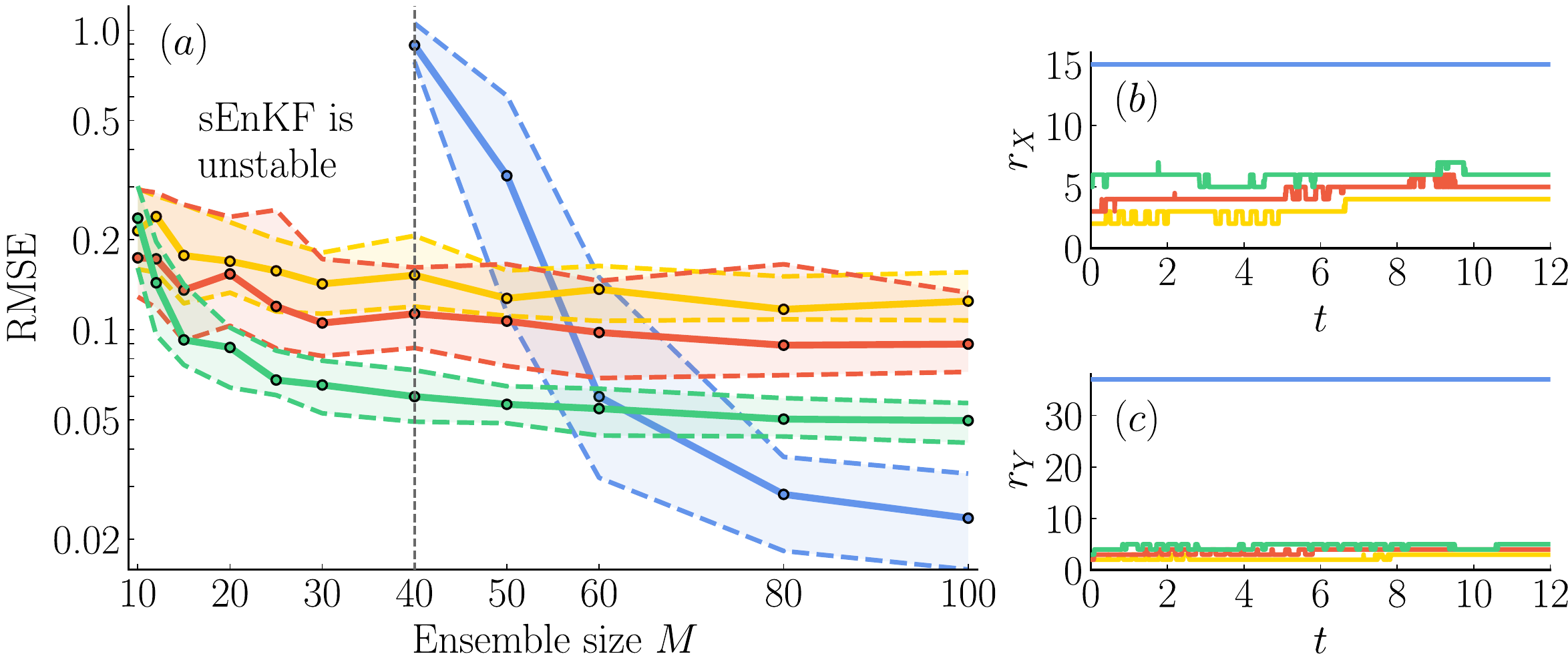}    \caption{Left column (a): Time-averaged evolution of the median RMSE with the ensemble size $M$ (computed over $50$ realizations) with the \senkf{} (blue), and with the \lrenkf{} for different ratios of the cumulative energy: $85\%$ (yellow), $95\%$ (orange), and $99\%$ (green). The tracking performance of the \senkf{} is unstable for $M<40$. Dashed lines depict the $25\%$ and $75\%$ quantiles. Right column [(b)-(c)]: Time-history of the median ranks $\rx$ and $\ry$ of the \lrenkf{} for $M = 30$ (computed over $50$ realizations) for different ratios of the cumulative energy of $\Cx$, $\Cy$: $85\%$ (yellow), $95\%$ (orange), and $99\%$ (green). The dimension of the state and observation spaces, namely $n$ and $d$, are depicted for comparison in blue.}
    \label{fig:vortex_rmserank}
\end{figure}

Next, we compare the performance of the \lrenkf{} with the  stochastic \enkf{} (\senkf{}).  The comparison of these two filters is natural, as the \lrenkf{} without dimension reduction (\ie for $\rx = n = 3 N$ and $\ry = d$) reverts to the \senkf{}. For a given ensemble size $M$, one could perform a parametric study to determine the ranks $\rx$ and  $\ry$ which give the best performance of the \lrenkf{}. We found, however, that using fixed ranks for all assimilation cycles is suboptimal as we face one of two possible scenarios: either the ranks are too large, leading to additional variance, or the ranks are too small, leading to additional bias. Furthermore, we expect the dimension of the informative subspaces may vary over time as the flow field evolves.  Instead, we select the ranks $\rx$ and $\ry$ adaptively at each time step based on a predetermined ratio of the cumulative energy of the Gramians $\Cx$ and $\Cy$.

The performance of each filter is assessed using the \textit{root mean squared error (RMSE)} for tracking the true state $\state_k^\star$. The best estimate of the true state at a given assimilation time is given by the mean of the analysis ensemble at this step, that we denote by $\overline{\state}^a_k$. We define the RMSE at one time between the true state $\state^\star_k$ and the estimate $\overline{\state}_k^a$ as $\rmse_k = ||\state_k^\star - \overline{\state}^a_k||_{2}/\sqrt{3  N}$,  where $N$ is the number of point vortices. The mean RMSE of each filter is computed over the time interval $[8, 12]$ (\ie the last 4000 assimilation steps), to remove any influence of the initial conditions and to ensure empirical stationarity of the filters. The first $8000$ assimilation steps are called the spin-up phase and are discarded. The uncertainty in the RMSE is quantified by the $5\%, 25\%, 75\%$ and $95\%$ quantiles of the time-averaged RMSE over $50$ realizations of the same experiment for different initial ensembles. The total computation time for one assimilation step of the \senkf{} and the \lrenkf{} with $M=50$ are $2.3$ ms and $8.0$ ms, respectively.
Given that the \lrenkf{} achieves a lower RMSE for small $M$, a fairer comparison is to determine the ensemble size needed by these two filters to achieve a tolerated tracking error. For instance, the \senkf{} needs $M=60$ ensemble members for a median RMSE of $0.07$, while the \lrenkf{} with a $99\%$ energy ratio only needs $M = 20$ ensemble members. Thus, the performances of the filters are fairly similar given that their computation times scale linearly with the ensemble size. We stress that the reported computation time of the \lrenkf{} corresponds to the worst case scenario where the entire Jacobian of the observation operator is evaluated at all samples to form the Gramians $\Cx$ and $\Cy$. We provide directions to reduce the computational cost of the \lrenkf{} in the conclusion.

 Fig.~\ref{fig:vortex_rmserank}(a) shows the time-averaged RMSE for the \lrenkf{} and the \senkf{}. We assess the performance of the \lrenkf{} for $85\%$, $95\%$, and $99\%$ of the normalized cumulative energy of $\Cx$ and $\Cy$. For $M<40$, the \senkf{} is unstable, while the \lrenkf{}  accurately estimates the true state, even for $M = 10$.  For $M \in [40,60]$ , the RMSE of the \senkf{} significantly increases as $M$ decreases. For $M = 40$, the median RMSE of the \senkf{} is $0.9$, while the median RMSEs of the \lrenkf{} with the different energy ratios are smaller than $0.16$. For large $M$, the RMSE of the \lrenkf{} decreases with the energy ratio, but remains larger than the RMSE of the \senkf{}. The proposed low-rank approximation of the Kalman gain can lead to a small bias for large ensemble sizes. The regularization technique proposed in this work, however, is designed for the small ensemble size regime; one can always increase the energy ratio (beyond $99\%$) to recover the performance of the \senkf{} for large ensemble sizes. The RMSE of the \lrenkf{} with the different energy ratios decreases for $M \in [10, 40]$, then plateaus. In the small ensemble size regime, adding samples improves the estimate of the dominant directions (see the Davis--Kahan theorem~\ref{thm:davis}), leading to a reduction in variance. As we continue to increase the number of samples for a given number of dimensions to estimate, the variance becomes smaller than the bias from using a truncated basis to approximate the Kalman gain. For $M = 10$, the \lrenkf{} with $85\%$ energy ratio performs better than the \lrenkf{} with $99\%$ energy ratio. For $M = 10$, the ranks needed to achieve $99\%$ is  large, leading to additional variance. On the other hand, $85\%$ energy requires a much smaller number of entries to estimate, leading to less variance compared to the bias of this approximation. We recommend choosing the target energy ratio based on the available ensemble size.

Fig.~\ref{fig:vortex_rmserank}(b)-(c) show the history of the median ranks $\rx$ and $\ry$ (computed over $50$ realizations) for the different ratios of the cumulative energy for $M = 30$. The ranks plateau over the time window and remain small compared to the dimension of the state and observation spaces. For the \lrenkf{} with $99\%$ of the cumulative energy, $\rx, \ry$ are smaller than $8, 6$, respectively. Fig.~\ref{fig:example2_trajectories} (b) shows one estimate of the trajectories of the vortices over the time interval $[0, 12]$. The results are obtained from the \lrenkf{} with $M = 50$ with the ranks $\rx$ and $\ry$ set to  capture $99\%$ of the energy spectra. We quantify the uncertainty with the $5\%$ and $95\%$ quantiles of the posterior for the point vortex positions. We observe an excellent agreement with the true trajectories with a time-averaged RMSE of $0.038$.

\begin{figure}
    \centering
    \includegraphics[width =1.0\linewidth]{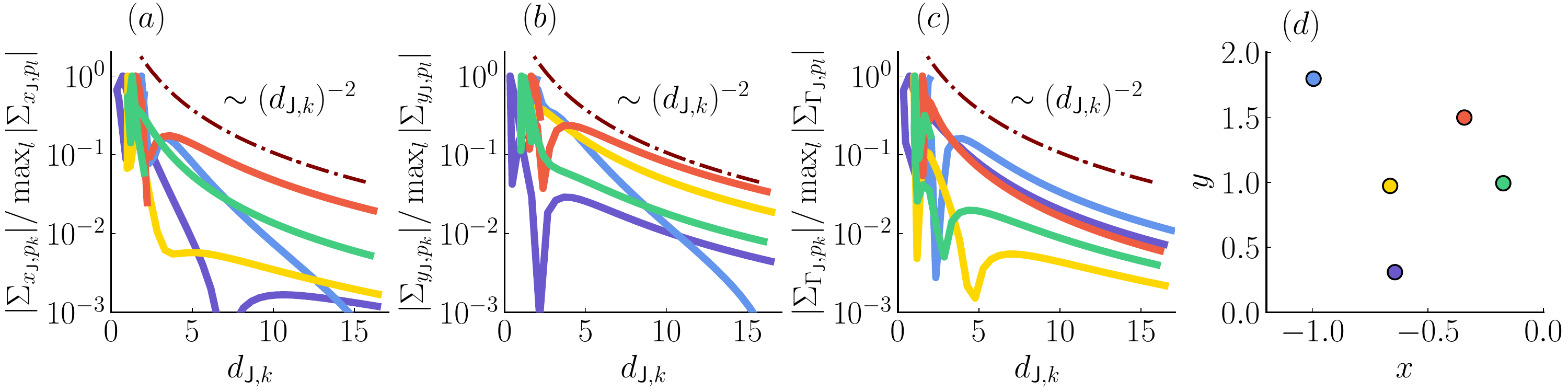}
    \caption{Panel (a), (b), (c) depicts the magnitude of the empirical cross-covariance between the $x$, $y$ coordinate and strength of the $\Jidx$ point vortex with the $k$th pressure observation at $t = 1.0$, respectively. The curves are plotted against the distance $d_{\Jidx, k} = |z_\Jidx - \zp_k|$, where $z_\Jidx = x_\Jidx + i y_\Jidx$ is the (complex) position of the $\Jidx$th point vortex, and $\zp_k$ is the (complex) location of the $k$th pressure sensor.  The magnitude of the cross-covariances are normalized by the maximum cross-covariance (in magnitude) between the $x$, $y$ coordinate or strength of the $\Jidx$ point vortex with the different pressure observation, respectively. The dashed grey vertical line depicts the $x$ component of the position of the centroid of the vortices. Red dashed and dotted curve depicts the algebraic decay $O(1/d_{\mathsf{J},k}^2)$. (d) Mean position of the different point vortices at $t = 1.0$. The same color is used to depict the properties of a point vortex on the panels (a)-(d). The results are obtained from a run of the \senkf{} for $M = 1000$.}
    \label{fig:example2_crosscorrelation}
\end{figure}

In Section~\ref{subsec:lowrank}, we argue that distance localization schemes can be harmful to regularize inference problems with elliptic observation models, as studied herein. We conclude this section by providing an \textit{a posteriori} justification based on the results of the \senkf{} run with a large ensemble ($M = 1000$). Fig.~\ref{fig:example2_crosscorrelation} (a)-(c) shows the  magnitude of the cross-covariance between the positions and strengths of the different point vortices and the pressure observations at $t = 1.0$. We notice that the cross-covariance entries decay very quickly on a small support about each point vortex. Past a certain radius, however, the cross-covariance entries quickly increase, and beyond this point, they only decay algebraically. More precisely, we observe a decay as the inverse square of the distance between the point vortex and the pressure observation. This decay rate is expected from the pressure field derived in~\eqref{eqn:pressuresolution}. In Fig.~\ref{fig:example2_crosscorrelation}(c), the covariance between the circulation of three point vortices and pressure observations at a distance of $5$ and $10$ units away is still about $20\%$ and $10\%$ of its maximal value, respectively. This algebraic decay clearly violates the assumption of rapidly decaying correlations of distance localization schemes. The cross-covariance of the components of the position and the strength for the different point vortices can have different algebraic decay rates as a function of the distance. Even for a particular vortex, the variations of the cross-covariance for different components of the position and the strength are drastically different. In unreported results, we observe that the decay rate of the correlations also varies over time as the vortices evolve. In the best case, this suggests that multiple different localization radii (i.e., three for the state and one for the data)  are needed in principle and would have to be tuned independently, which is typically impractical. These different facts support our \textit{a priori} hypothesis that distance localization is not uniformly well suited to regularize filtering problems with elliptic observation operators. 
Moreover, while distance localization can be successful in regularizing the \senkf{} for some problems, its underlying assumptions are clearly violated by the algebraic decay in this example. Thus, the application of distance localization should be carefully considered. In unreported results that motivated this work \footnote{See \url{https://github.com/mleprovost/LocalizedVortex.jl.git} for the numerical results of the localized EnKF.}, distance localization was detrimental and worsened the  performance of the \enkf{} in some aerodynamic estimation problems; see \cite{leprovost2021ensemble} for details on the setting.


\subsection{Inference of a vortex patch advected along a wall from pressure observations. \label{subsec:vortexpatch}}

\begin{figure}
    \vskip -0.4cm
    \centering
    \includegraphics[width = 0.85\linewidth]{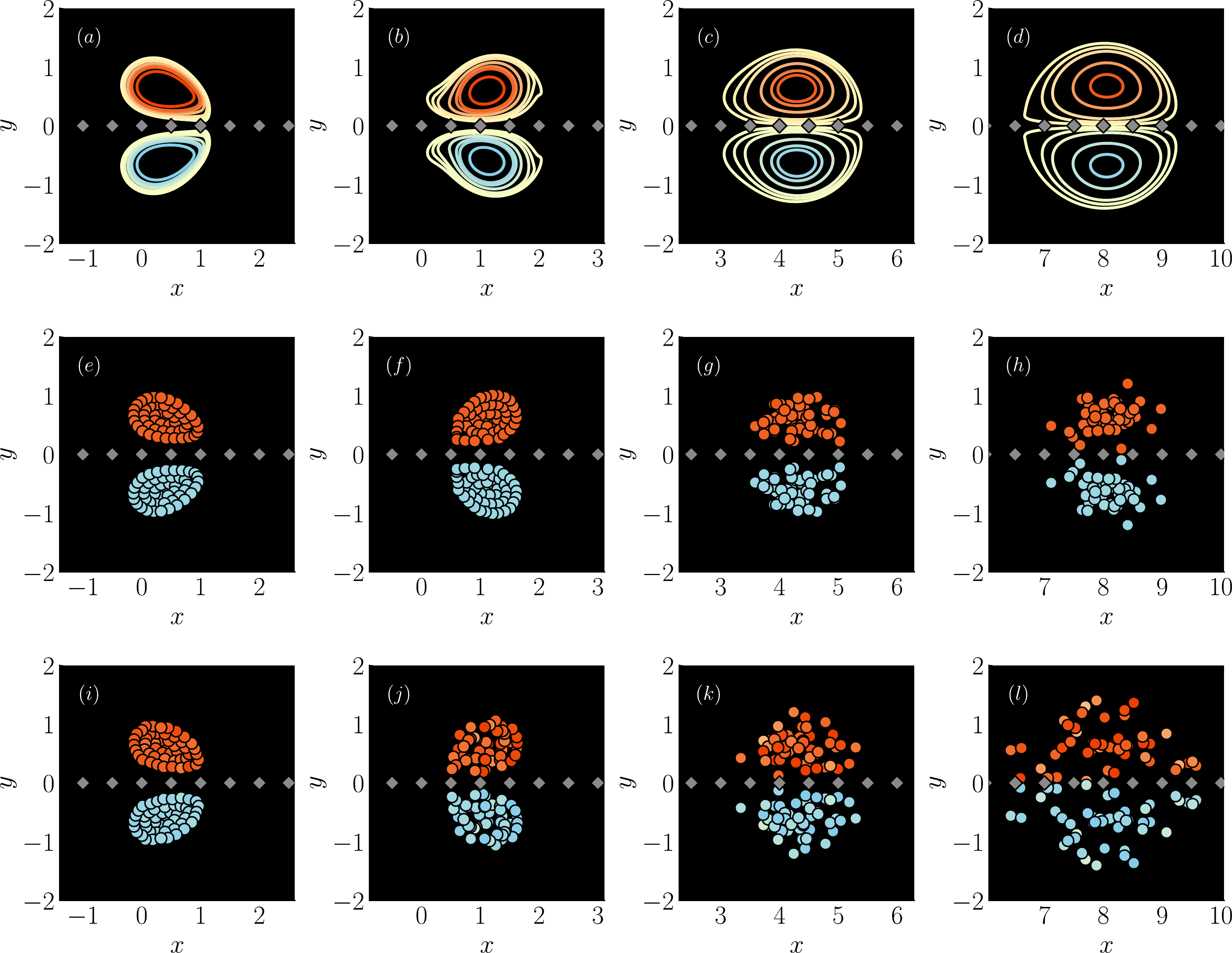}
    \caption{Snapshots of the vorticity distribution at $t = 0.5$ (left column), $t = 1.5$ (second column), $t = 6.0$ (third column), and $t = 12.0$, for a vortex patch and its image, predicted from [(a)-(d)] high-fidelity numerical simulation at Reynolds number $1000$, [(e)-(h)] inviscid vortex model with \lrenkf{} with $M = 30$, and [(i)-(l)] inviscid vortex model with \senkf{} with $M = 30$.  The ranks $\rx$ and $\ry$ of the \lrenkf{} are set to capture $99\%$ of the normalized cumulative energy of $\Cx$ and $\Cy$, respectively. Pressure sensors are depicted with grey diamonds. Orange denote positive vorticity, while blue vorticity denotes negative vorticity.}
    \label{fig:vortexpatch_vortex}
\end{figure}

In this third example, we infer the properties of a circular vortex patch advected along the $x$ axis. A circular vortex patch $\mathcal{V}$ is a disk of uniform vorticity $\omega_0$. As in~\ref{subsec:pointvortex}, the vortex patch is constrained to advect along the $x$ axis by the method of images~\cite{eldredgebook}. The total circulation of the vortex patch $\mathcal{V}$, denoted $\Gamma_\mathcal{V}$, is given by $\Gamma_\mathcal{V} = \int_{\state \in \mathcal{V}}  \omega_0 \; \mathrm{d}S(\state)$ and the vorticity centroid, denoted $\state_{\mathcal{V}}$, is given by  $\state_{\mathcal{V}} = {\Gamma_{\mathcal{V}}} \int_{\state \in \mathcal{V}}  \state \omega_0 \; \mathrm{d}S(\state)$. The problem is parameterized by the ratio of the radius $r_{\mathcal{V}}$ of the vortex patch to the distance $d_{\mathcal{V}}$ between the centroid of the vortex patch and its image centroid. 

We discretize each vortex patch with a collection of regularized point vortices; further details are given in~\cite{eldredgebook}. In this study, the vortex patch is discretized by $N_r = 4$ concentric rings of blobs, leading to a subdivision of the vortex patch into $N_{pv} = 1 + 4 N_r (N_{r -1}) = 49$. The initial position of the $\Jidx$th point vortex is generated randomly with the form $z_{\Jidx} + \rho_r \exp(i \theta)$, where $z_{\Jidx}$ denotes the nominal position of the $\Jidx$th vortex. $\rho_r$ is drawn from $\N(0.0, \sigma_r^2)$, where $\sigma_r$ corresponds to $10\%$ of the radius between two concentric rings of point vortices, and $\theta$ is drawn from the uniform distribution on $[0, \; \pi]$. The initial circulation of the point vortices is drawn from $\N(\Gamma_{\mathcal{V}}/N_{pv}, 10^{-4})$. For a blob at $z_\Jidx$ with circulation $\Gamma_\Jidx$ in the vortex patch $\mathcal{V}$, there is an image blob located at $\overline{z_{\Jidx}}$ with circulation $-\Gamma_\Jidx$. The state is defined, as in the previous example, by the positions and strengths of the point vortices in the upper half-plane.  For $N_{pv}$ blobs, the state dimension is $n = 3 N_{pv} = 147$. We use a forward Euler model scheme with time step $\Delta t = 5.0 \times  10^{-3}$. The vortex patches are evolved over the time interval $[0, 12]$. The observation vector $\meas$ consists of pressure observations collected at $d = 24$ locations with interspace $0.5$ along the segment $[-1.5,10.0]$ of the $x$ axis.  The observation noise is Gaussian with zero mean and covariance $4 \times 10^{-2}$ (corresponding to a ratio of peak pressure amplitude to standard deviation of the noise equal to $10$).

In this example, the dynamical model that generated the observations is given by solving the incompressible Navier-Stokes equations at a Reynolds number $\Gamma_\mathcal{V}/\nu = 1000$, where $\nu$ is the kinematic viscosity of the fluid. The true solution contains viscous effects, which are not explicitly included in the forecast model we use for inference. The initial condition is constructed by discretizing the two vortex patches on a uniform Cartesian grid with grid spacing $\Delta x = 0.01$ for the same geometry (\ie $r_\mathcal{V}, \state_\mathcal{V},$ and $d_\mathcal{V}$ are identical) and the same total circulation $\Gamma_\mathcal{V}$. We use a high-fidelity Navier Stokes solver with lattice Green function (LGF)~\cite{taira2007immersed, liska2017fast}. The true pressure observations are generated by inverting the pressure Poisson equation~\eqref{eqn:pressurePoisson} with LGF. Panels (a)-(d) of Fig.~\ref{fig:vortexpatch_vortex} depict the truth vorticity field at $t = 0.5, 1.5, 6.0$ and $12.0$. 

To the best of our knowledge, there is no straightforward way to compare the discrete vorticity distribution generated by the ensemble filter with the true continuous vorticity distribution obtained from the Navier-Stokes solver. Instead, we compare the pressure distribution along the $x$ axis from the high-fidelity simulation with the evaluation of the observation model~\eqref{eqn:obs} at the posterior ensemble generated by the  \lrenkf{}, or the \senkf{} (\ie posterior predictive samples for the pressure distribution).                                                                                                      The performance of the \lrenkf{} is assessed for $85\%, 95\%$ and $99\%$ of the normalized cumulative energy for $\Cx$ and $\Cy$. We assess the performance of the filter on the segment $[-1.5, 11.0]$ with a finer interspace distance of $0.01$. We define the median-squared-error (MSE) at time $k$ between the true pressure distribution $\meas^\star_k$, and the median posterior pressure distribution $\text{median}\left({\meas}^a_k\right)$ (\ie the median pressure distribution for the posterior ensemble at time $k$) as $\text{MSE}_k = ||\meas^\star_k - \text{median}\left({\meas}^a_k\right)||_2/\text{dim}(\meas^\star_k)$, where $\text{dim}(\BB{v})$ denotes the dimension of a vector $\BB{v}$. Due to the singular nature of the pressure field, we found that the MSE is a more robust estimator of the performance of each filter than the RMSE due to the presence of outliers in the ensemble estimate of the \senkf{}. The spin-up phase of the filters is $[0.0, 8.0]$.  The reported MSE is time-averaged over the remaining time interval $[8.0, 12.0]$. The uncertainty in the MSE is quantified by the $5\%, 25\%, 75\%$ and $95\%$ quantiles of the MSE over $50$ realizations of the same experiment with different samples for the initial condition. The computation time for one assimilation step of the \senkf{} and the \lrenkf{} with $M=50$ are $9.8$ ms and $42.5$ ms, respectively.
To achieve a median MSE of at most $0.04$, the \senkf{} needs $M= 30$ ensemble members, while the \lrenkf{} with the different energy ratios studied only needs $M = 12$ ensemble members.

\begin{figure}
    \includegraphics[width = \linewidth]{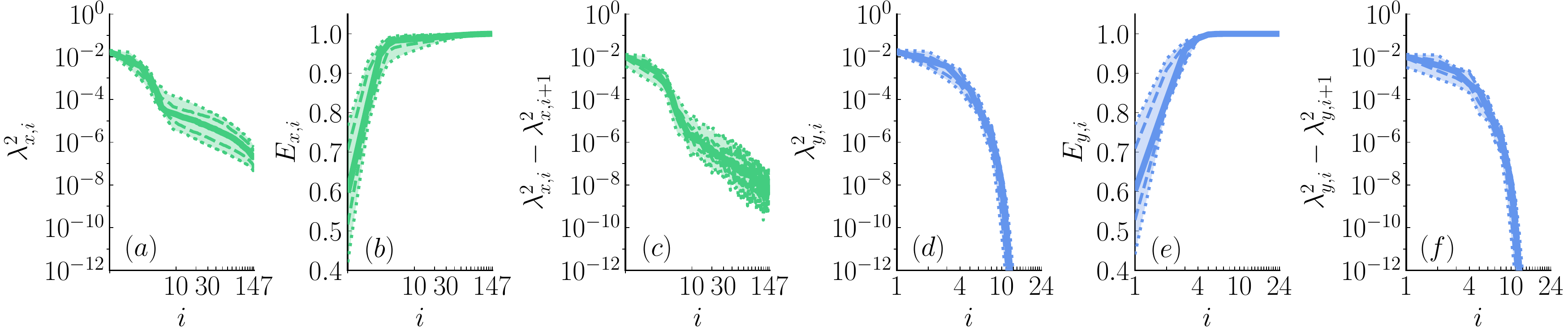}
    \caption{Median spectrum of the state Gramian $\Cx$ (left panels (a)-(c)) and observation Gramian $\Cy$ (right panels (d)-(f)) over the time interval $[0, 12]$. Statistics are obtained from a run of the \senkf{} with $M = 1000$. Panels [(a), (d)]: Median eigenvalues of $\Cx$ and $\Cy$. Panels [(b), (e)]: Median normalized cumulative energy $E_i = \sum_{j=1}^i\lambda^2_j/\sum_{j}\lambda^2_j$ of $\Cx$ and $\Cy$. Panels [(c), (f)]: Median spectral gap $\lambda^2_i - \lambda^2_{i+1}$ of $\Cx$ and  $\Cy$. The abscissa axis is in log scale. Dashed lines depict the $25\%$ and $75\%$ quantiles. Dotted lines depict the $5\%$ and $95\%$ quantiles.}
    \label{fig:vortexpatch_energyspectrum}
\end{figure}

Fig.~\ref{fig:vortexpatch_energyspectrum} shows the median spectra of $\Cx$ and $\Cy$ over the entire time interval. The statistics are obtained from a run of the \senkf{} for a large ensemble ($M = 1000$). As in the previous example, the inference problem possesses low-rank informative structure. The median ranks $\rx$ needed to capture $80\%, 90\%, 95\%,$ and $99\%$ of the cumulative energy of $\Cx$ are $2, 3, 4,$ and $16$, respectively. Similarly, the median ranks $\ry$ to capture $80\%, 90\%,$ and $99\%$ of the cumulative energy of $\Cy$ are $2, 3,$ and $5$, respectively. These ranks are small compared to the state dimension $n = 147$, and the observation dimension $d = 24$. 

\begin{figure}
    \centering
    \includegraphics[width = 0.8\linewidth]{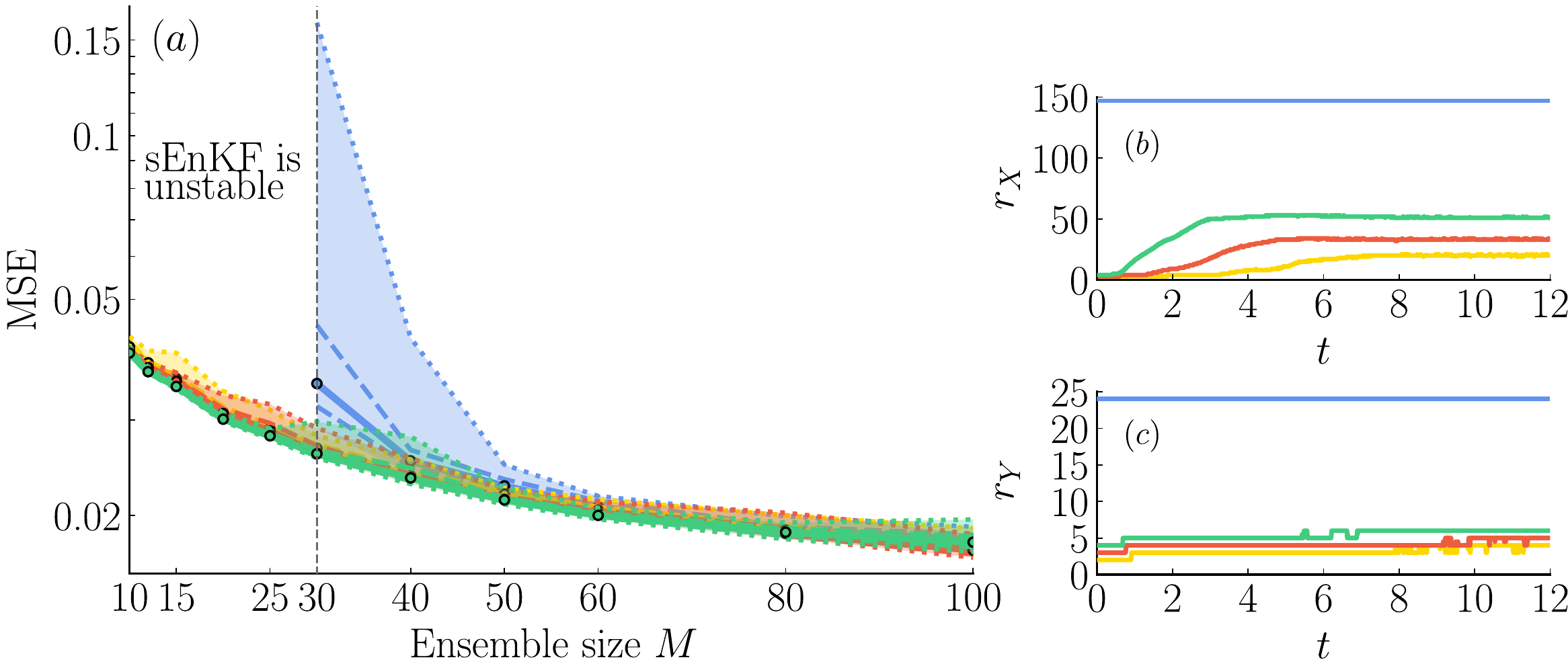}    \caption{Left column (a): Time-averaged evolution of the median MSE of the posterior pressure with the ensemble size $M$ (computed over $50$ realizations) with the \senkf{} (blue), and with the \lrenkf{} for different ratios of the cumulative energy: $85\%$ (yellow), $95\%$ (orange), and $99\%$ (green). The \senkf{} is found unstable for $M<30$. Dashed lines depict the $25\%$ and $75\%$ quantiles. Dotted lines depict the $5\%$ and $95\%$ quantiles. Right column [(b)-(c)]: Time-history of the median ranks $\rx$ and $\ry$ of the \lrenkf{} for $M = 20$ (computed over $50$ realizations) for different ratios of the cumulative energy of $\Cx$, $\Cy$: $85\%$ (yellow), $95\%$ (orange), and $99\%$ (green). The dimension of the state and observation spaces, namely $n$ and $d$, are depicted for comparison in blue.}
    \label{fig:vortexpatch_mserank}
\end{figure}

 Fig.~\ref{fig:vortexpatch_mserank} (a) reports the evolution of the median MSE (computed over $50$ realizations) for the \senkf{} and the \lrenkf{} with the ensemble size $M$. We assess the performance of the \lrenkf{} for $85\%, 95\%,$ and $99\%$ of the normalized cumulative energy of $\Cx$ and $\Cy$.  We recall that the \lrenkf{} reverts to the \senkf{} when the dimensions are not reduced. For $M  \geq 60$, there is no significant difference in the MSE of the \senkf{} and the \lrenkf{} (for the different energy ratios). For $M \in [30, 50 ]$, the MSE of the \senkf{} significantly increases as $M$ decreases. Over this interval, the MSE of the \lrenkf{} shows no major variation.  Overall, capturing $85\%$ of the cumulative energy of $\Cx$ and $\Cy$ is sufficient to yield stable inference results. For $M<30$, the MSE of the \senkf{} diverges, while the \lrenkf{} leads to a reasonable estimate of the pressure distribution for ensemble size as small as $10$. This clearly demonstrates the benefit of our regularization for the \enkf{}. Fig.~\ref{fig:vortexpatch_mserank}(b)-(c) shows the time history of the median ranks $\rx$ and $\ry$ required to capture at least $85, 95, 99\%$ of the cumulative energy (computed over $50$ realizations) for $M = 20$. The rank $\rx$ initially increases with time, until it reaches a plateau at about $t= 3.0-7.0$ depending on the energy ratio. The increase of the rank $\rx$ can be related to the growing role of viscosity in the true pressure response. The rank $\ry$ remains close to $5$ over time for the different energy ratios.

Fig.~\ref{fig:vortexpatch_vortex} compares the vorticity distribution at four times, $t = 0.5, 1.5, 6.0,$ and  $12.0$, from the truth with the distribution of vortex elements of the posterior mean of the \lrenkf{} and the \senkf{}. The ensemble estimates are obtained from one realization of each filter for $M = 30$.  The ranks of the \lrenkf{} are set to capture at least $99\%$ of the normalized cumulative energy. The large-scale vortex features are well captured by the inviscid vortex model, despite the obvious absence of the viscous effects in our flow representation. The state estimate is informed of the viscous effects through the assimilation of the pressure observations ~\cite{leprovost2021ensemble}. Overall, the \lrenkf{} provides a more physically consistent estimate of the vorticity distribution than the \senkf{}. With a limited ensemble size, the \senkf{} cannot distinguish between physical and spurious long-range correlations between the pressure observations and the vortices. As a result, the \senkf{} inconsistently displaces the vortices. At $t = 12.0$, the vorticity distribution of the \lrenkf{} contains a dense core of vortices, with a few vortex satellites to capture the viscous diffusion of the vortex patch. However, the \senkf{} poorly estimates the truth vorticity distribution for $t > 6.0$. The spatial structure of the vortices is lost, and they occupy a much larger support than the true vorticity distribution.

\begin{figure}
    \centering
    \includegraphics[width = \linewidth]{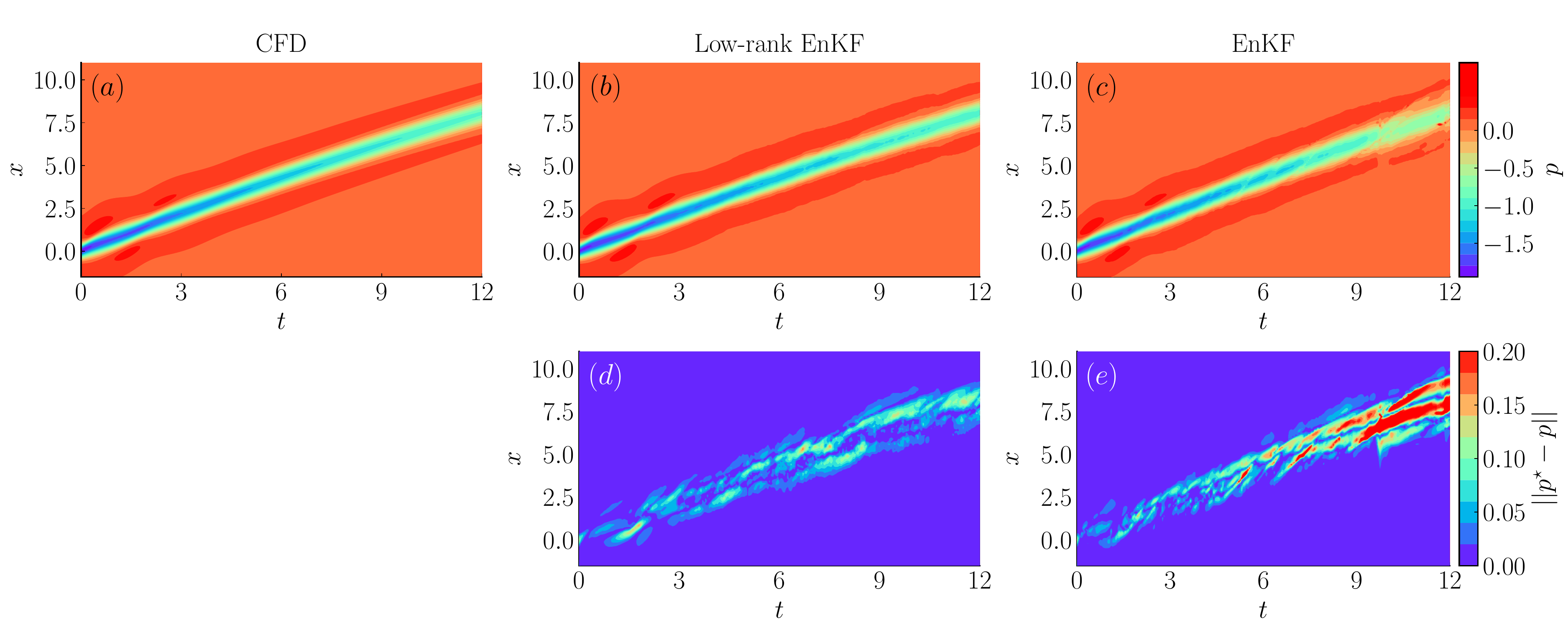}
    \caption{Top line [(a), (b), (c)]: Spatiotemporal map of the pressure induced by a pair of two co-rotating vortex patches from (a) high-fidelity numerical simulation at Reynolds number $1000$, and the mean posterior of an inviscid vortex model for one realization of the \lrenkf{} (b), and  the \senkf{} (c) with $M = 30$. Bottom line [(d), (e)]: Spatiotemporal magnitude of the error between the truth pressure and the mean posterior pressure for one realization of the \lrenkf{} (d), and the \senkf{} (e) with $M = 30$. The ranks $\rx$ and $\ry$ of the \lrenkf{} are set to capture $99\%$ of the normalized cumulative energy of $\Cx$ and $\Cy$, respectively. The same color levels are used for all the top panels, and another set for the bottom panels.}
    \label{fig:vortexpatch_pressure}
\end{figure}

Fig.~\ref{fig:vortexpatch_pressure} compares the history of the pressure from the truth with the posterior mean estimate of the \lrenkf{} and the \senkf{}  for the same realization of the filters. Over time, the pair of vortex patches is advected along the $x$ axis. Concomitant to the diffusion of the vortex patches, the peak pressure (in magnitude) decreases over time. The true pressure is globally well approximated by the two filters. As we pointed in the previous paragraph, the \senkf{} creates spurious displacements of the vortices over time, leading to a growing error in the posterior predictive pressure at later times (the red regions in Fig.~\ref{fig:vortexpatch_pressure}(e) for $t > 6.0$).

\section{Conclusion \label{sec:conclusion}}
In this work, we presented a new regularization technique for the \enkf{} with elliptic observation models. In these problems, the strength of interactions between states (\ie sources) and the observations (\ie targets) decays slowly (typically algebraically) with physical distance. As a result, distance localization  should be considered with care, as we cannot dissociate true/physical long-range correlations from the spurious long-range correlations induced by limited ensemble sizes. Instead, we observe that it is sufficient to update a limited number of directions of the state space using a low-dimensional projection of the observations. We propose to leverage this structure to regularize the \enkf{}. This structure yields a low-rank factorization of the Kalman gain in  linear--Gaussian observation models, which is computed from the SVD of a whitened observation matrix. We extended this factorization to  nonlinear--Gaussian observation models with the help of two tools. First, we proposed a methodology to identify the informative directions of the state and observation spaces for a nonlinear observation model. These directions are obtained as the leading eigenvectors of the state Gramian $\Cx$ and observation Gramian $\Cy$, which are based on the Jacobian of the observation operator $\obs$. These eigenvectors revert to the singular vectors of the whitened observation model (up to a sign) in the linear--Gaussian case. We show how to use these informative directions to define a prior-to-posterior transformation that updates a low-dimensional subspace of the states based on a projection of the observations. This transformation requires estimating a lower-dimensional Kalman gain in the span of the informative directions, thereby resulting in estimators with \textit{lower variance} than the \senkf{}.  Furthermore, for elliptic observation operators, the spectra of $\Cx$ and $\Cy$ are rapidly decaying. This makes the bias from low-rank approximations small for any budget of samples. 
Setting the ranks $\rx$ and $\ry$ a priori is not desired, as the dimension of the informative subspace may vary over time (see Examples~\ref{subsec:pointvortex},~\ref{subsec:vortexpatch}). Instead, we showed that the ranks $\rx$ and $\ry$ can be adaptively selected to capture a given ratio of the cumulative energy of $\Cx$ and $\Cy$. We stress that these reduced coordinates are not limited to the estimation of low-dimensional linear transformations. Le Provost et al.~\cite{leprovost2021low} used this informative subspace to identify parsimonious and interpretable \textit{nonlinear} prior-to-posterior transformations for aerodynamic applications \cite{spantini2019coupling, baptista2020adaptive}.

In Example \ref{subsec:CxCymultipole}, we showed that the leading eigenvectors of $\Cx$ and $\Cy$ are reminiscent of the modes from multipole expansions. Nonetheless, the proposed methodology is more general, as it only relies on the Jacobian of the observation operator and works even when observations are taken near the sources. In Example \ref{subsec:pointvortex}, the first eigenvectors of $\Cx$ form a hierarchical basis of discrete stencils to extract information from the pressure observations. We assessed the \lrenkf{}  and the \senkf{} on two problems where we sought to estimate the positions and strengths of point vortices from pressure observations. In the second example, the pressure observations were obtained from a high-fidelity simulation of the Navier-Stokes equations. In these two examples, we showed that the \lrenkf{} significantly improved the posterior estimate of the \senkf{} by yielding lower RMSE for smaller ensemble sizes.

One limitation of this work is the computational cost of forming the Gramians $\Cx$ and $\Cy$. Fortunately, two key ideas can be used to reduce this cost. First, we do not need the entire set of eigenpairs of $\Cx$ and $\Cy$. The \lrenkf{} only requires a basis for the leading eigenspaces of $\Cx$ and $\Cy$. Given rapid spectral decay of these Gramians, the subspaces spanned by their leading eigenvectors can be estimated with a relatively limited number of samples. Therefore, it may not be necessary to evaluate the Jacobian of the observation operator for every member of the ensemble; instead one could consider a subset of the ensemble, selected by random subsampling or perhaps clustering.

Second, computation of the full Jacobian matrix for each sample is not necessary. By leveraging ideas from randomized numerical linear algebra~\cite{halko2011finding, eftekhari2016randomized}, one can construct accurate estimates of the leading eigenspaces of $\Cx$ and $\Cy$ using only matrix-vector products.

We also emphasize that the proposed regularization of the \enkf{} is by no means limited to potential flow problems, nor fluid mechanics more generally. Indeed, elliptic observation models are found across a spectrum of data assimilation problems in engineering and science: geoscience, electromagnetism, and heat transfer, to name a few. For all these problems, regularization of the \enkf{} has usually been \textit{ad hoc}~\cite{bocher2018ensemble, lang2017data}. The principled regularization of the \enkf{} described in this paper is applicable to any differentiable observation model, without any tuning required from the user---the only free parameters being the ranks $\rx$ and $\ry$ which can be set adaptively, as in the examples we presented. 

Finally, we comment on the link between our methods and \textit{inverse problems}. While the focus of this paper was on \textit{filtering} problems and their solution using ensemble methods, every assimilation step with an elliptic observation operator is essentially a Bayesian inverse problem, as noted in the Introduction. The inverse problems viewpoint inspires the regularization approaches we develop, and in fact these regularization methods are immediately applicable to ensemble algorithms \cite{garbuno2020interacting, reich2021fokker} for a wide range of Bayesian inverse problems, as well as to other solution algorithms that could benefit from joint dimension reduction of the data and parameters \cite{kovachki2020conditional}. In other words, our developments are not restricted to filtering. That said, the specific examples and problem settings that we consider in this paper, as well as the performance metrics and comparisons we use for evaluation (e.g., RMSE, small ensemble sizes), are representative of filtering applications. Evaluation of these methods for the typical settings of Bayesian inverse problems is left to future work.

\vskip6pt

\enlargethispage{20pt}

\dataccess{All the computational results are reproducible and code is available at: \url{https://github.com/mleprovost/LowRankVortex.jl.git}}

\aucontribute{All authors contributed to the calculations, analysis, and writing and revision of the manuscript. The final version of the manuscript was approved by all authors.}

\funding{MLP and JE gratefully acknowledge support of the Air Force Office of Scientific Research Grant No. FA9550-18-1-0440. MLP acknowledges support of the Dissertation Year Fellowship of the Graduate Division of the University of California, Los Angeles. RB and YM acknowledge support from the Department of Energy, Office of Advanced Scientific Computing Research, AEOLUS (Advances in Experimental design, Optimal control, and Learning for Uncertain complex Systems) center.}

\section*{\begin{center}Appendix \end{center}}
\begin{appendices}

\section{The transport equations for singularities\label{secSM:biotsavart}}
Here, we present the derivation of the transport equations for singularities in potential flow. We recall for reference the complex potential $F$ induced at a location $z \in \complex{}$ by a collection of $N$ singularities located at $\{z_1, \ldots, z_N\}$ with complex strengths $\{\strength_1, \ldots, \strength_N\}$ subject to a freestream velocity given in complex notation $\Wstream = U_{\infty} - i V_{\infty}$.
\begin{equation}
\label{eqnSM:complexpotential}
    F(z) = \sum_{\Jidx = 1}^N \frac{S_\Jidx}{2\pi} \log(z - z_\Jidx) + \Wstream z.
\end{equation}
The derivative of the complex potential is the complex velocity $w$ given by:
\begin{equation}
\label{eqnSM:vel_complexnotation}
    w(z, t) = \diff{F}{z} = \sum_{\Jidx = 1}^N \frac{S_\Jidx}{2\pi} \frac{1}{z - z_\Jidx} + W_{\infty}.
\end{equation}
For a velocity field given in vector notation by $\vel(\state, t) = (u(\state, t), v(\state, t))$, we stress the classical conjugation (\ie minus sign) in the definition of the equivalent complex velocity $w(z, t) = u(z, t) - i v(z, t)$. Using the induced velocity \eqref{eqnSM:vel_complexnotation}, we can write the transport equation for the positions and strengths of the singularities. By simple inspection of \eqref{eqnSM:vel_complexnotation}, the velocity field is singular at the locations of the singularities. Per Kirchhoff's law, a point singularity cannot induce velocity on itself \cite{eldredgebook}. Instead, a singularity is transported by the total induced velocity on itself minus its own contribution. The regularized velocity of the $\Jidx$th singularity is called the Kirchhoff velocity, denoted $\vel_{-\Jidx}$ in vector notation, or $w_{-\Jidx}$ in complex notation. The transport equations for the singularities follow: 
\begin{equation}
\label{eqnSM:biotsavart}
    \ddt{z_\Jidx}  = \overline{w_{-\Jidx}}(t) = \sum_{\Kidx = 1, \Kidx \neq \Jidx}^N \frac{\overline{S_\Jidx}}{2\pi} \frac{1}{\overline{z_\Jidx} - \overline{z_\Kidx}} + \overline{W_{\infty}}, \hspace{0.3cm}
    \ddt{S_\Jidx}   = 0, \hspace{0.3cm} \Jidx=1, \ldots, N.
\end{equation}

\section{The pressure Poisson equation and the Bernoulli equation \label{secSM:pressure}}
In this section, we present the derivation of the pressure Poisson equation for potential flows and the analytical expression for the pressure field obtained from the unsteady Bernoulli equation. The Euler equation and the associated divergence and curl conditions on the velocity field are given by: 
\begin{subequations}
\begin{equation}
\label{eqnSM:euler}
\rho \diffp{\vel}{t} + \rho \vel \cdot \nabla \vel  = -\nabla p ,   
\end{equation}
\begin{equation}
\label{eqnSM:divergence}
 \diver{\vel}(\position, t) = \sum_{\Jidx = 1}^N \flux_\Jidx \delta(\position - \position_\Jidx),
\end{equation}
\begin{equation}
\label{eqnSM:curl}
 \curl{\vel}(\position, t) = \sum_{\Jidx = 1}^N \circu_\Jidx \delta(\position - \position_\Jidx).  
\end{equation}
\end{subequations}
The pressure Poisson equation is formed by taking the divergence of the Euler equation \eqref{eqnSM:euler} and using the identity $\vel \cdot \nabla \vel  = \nabla(||\vel||^2/2) -\vel \times \vort$: 
\begin{equation}
\label{eqnSM:pressurePoissonnotsimplified}
    \lap\left(p + \frac{1}{2}\rho ||\vel||^2 \right) = \rho \diver\left(\vel \times \vort \right)  - \rho \diffp{\diver{\vel}}{t},
\end{equation}
where $\vort(\state, t)$ is the vorticity. Using \eqref{eqnSM:divergence} and \eqref{eqnSM:curl}, we obtain the pressure Poisson equation: 
\begin{equation}
\label{eqnSM:pressurePoisson}
    \lap\left(p(\position, t) + \frac{1}{2}\rho ||\vel(\position, t)||^2 \right) = \rho \sum_{\Jidx = 1}^N \left[\flux_{\Jidx} \vel_{-\Jidx}  \cdot \nabla \delta(\position - \position_{\Jidx}) - \circu_\Jidx \BB{e}_z \cdot (\vel_{-\Jidx} \times \nabla \delta(\position - \position_\Jidx)) \right].
\end{equation}

For a fixed evaluation point denoted $\positionp$ in vector notation or $\zp$ in complex notation, the unsteady Bernoulli equation reads \cite{eldredgebook}: 
\begin{equation}
\label{eqnSM:bernoulli}
p(\positionp, t) + \rho \frac{1}{2}||\vel(\positionp, t)||^2 + \rho \pddt{\pot(\positionp, t)} = B(t),
\end{equation}
where $B(t)$ is the Bernoulli constant. The second term of \eqref{eqnSM:bernoulli} is called the quadratic term, while the third term is called the unsteady term. Using the results derived in \ref{secSM:biotsavart}, we get
\begin{equation}
\label{eqnSM:pressuresolution}
    p(\zp, t) - B(t) =  -\frac{1}{2}\rho  \left| \sum_{\Jidx = 1}^N \frac{S_\Jidx}{2\pi} \frac{1}{\zp - z_\Jidx} + W_\infty\right|^2
     + \rho \re{\sum_{\Jidx = 1}^N \frac{S_\Jidx}{2\pi} \frac{1}{\zp - z_\Jidx} \overline{w_{-\Jidx}}(t)}.
\end{equation}

\section{\label{secSM:filtering}The filtering problem}

In this section, we give a basic outline of the filtering problem. We recall for reference the dynamical and observation models used in this work. The dynamical model is given by
\begin{equation}
\label{eqnSM:dynamic}
    \State_{k} = \dyn_k(\State_{k-1}) + \Noisedyn_k,
\end{equation}
where $\dyn_k\colon \real{n} \to \real{n}$ is a deterministic function and $\Noisedyn_k$ is an independent \textit{process noise}. The observation model is given by a nonlinear observation operator $\obs_k :\real{n} \to \real{d}$ corrupted by an additive observation noise $\Noiseobs_k$:
\begin{equation}
\label{eqnSM:obs}
    \Meas_k = \obs_k(\State_k) + \Noiseobs_k.
\end{equation}
In the \textit{filtering problem}, we aim to characterize the \textit{filtering distribution} $\pdf{\State_k \given \Meas_{1:k} = \meas_{1:k}}$ (also called the \textit{posterior} distribution), \ie the  probability distribution of the state at time $k$ conditioned on the realizations $\meas_1, \ldots, \meas_k$ of the observation process up to that time. To tackle this problem in high dimensions, we consider a Monte Carlo approximation of the filtering distribution. To do so, we recursively update a set of $M$ \textit{particles} $\{ \state^1, \ldots, \state^M \}$ that approximate the posterior distribution. This defines a generic class of methods known as ensemble filtering methods. At every time step, the recursion takes as input the particle approximation $\{ \state^1, \ldots, \state^M \}$ of the posterior distribution at time $k-1$ and seeks a particle approximation of the posterior  distribution at the next time. These methods recursively apply a two-step procedure: the \textit{forecast step} and the \textit{analysis step}. In the forecast step, each particle $\state^i$ is propagated in the dynamical equation~\eqref{eqnSM:dynamic} to form a Monte Carlo approximation of the forecast distributon $\pdf{\State_k \given \Meas_{1:k-1} = \meas_{1:k-1}}$. The analysis step treats the forecast as the prior and solves a Bayesian inference problem to condition on a new observation of the true system $\meas^\star_k$. Ensemble methods realize this conditioning by moving particles $\{\state^i\}_{i=1}^M$ that represent the prior/forecast distribution into new positions so that they form an empirical approximation of the posterior. We note that different ensemble filtering methods have a common forecast step but differ in the analysis step. Our treatment of the analysis step builds a transformation $\tmap_k$ that essentially maps samples from the prior distribution $\pdf{\State_k \given \Meas_{1:k} = \meas_{1:k-1}}$ to the posterior distribution $\pdf{\State_k \given \Meas_{1:k} = \meas_{1:k}}$ \cite{marzouk2016sampling, spantini2019coupling}. For a linear--Gaussian state-space model, Kalman~\cite{kalman1960new} derived in closed form an exact linear prior-to-posterior transformation for the analysis step. In more general settings, the prior-to-posterior transformation is necessarily nonlinear, but the linear transformation of the Kalman filter is still a foundation for algorithms such as the \textit{ensemble Kalman filter (EnKF)} introduced by Evensen \cite{evensen1994sequential}. In particular, the \enkf \ and other ensemble filtering methods estimate a linear analysis map $\tmap_k$ from samples. 
%
%
In order to simplify notation, we omit the time dependence subscripts of the variables in the rest of this Supplementary Material, since the analysis step can be treated as a \textit{static} Bayesian inference problem.

\section{\label{secSM:linearlowrank}Low-rank assimilation for the linear--Gaussian case}

In this section, we present a detailed derivation of the factorization of the Kalman gain for a linear--Gaussian observation model, recalled for reference:
  \begin{equation}
\label{eqnSM:linobs}
\Meas = \Obs \State + \Noiseobs,
\end{equation}
where the state is given by $\State \sim \N(\zero{}, \cov{\State})$ and the observational error is given by $\Noiseobs \sim \N(\zero{}, \cov{\Noiseobs})$ where $\Noiseobs$ is independent of $\State$. The whitened state and noise observation variables are defined as $\tilde{\State} = \cov{\State}^{-1/2} \State \in \real{n}$ and $\tilde{\Noiseobs} = \cov{\Noiseobs}^{-1/2}\Noiseobs \in \real{d}$. In the whitened space, the observation model becomes: 
\begin{equation}
\label{eqnSM:whitenedlik}
\tilde{\Meas} = \cov{\Noiseobs}^{-1/2} \Meas = \tilde{\Obs} \tilde{\State} + \tilde{\Noiseobs},
\end{equation}
where $\tilde{\Obs} = \cov{\Noiseobs}^{-1/2} \Obs \cov{\State}^{1/2} \in \real{d \times n}$ is the whitened observation matrix, whose singular value decomposition (SVD) reads: 
\begin{equation}
    \tilde{\Obs} = \BB{U} \BB{\Lambda} \BB{V}^\top,
\end{equation}
where $\BB{U} \in \real{d \times d}$ and $\BB{V} \in \real{n \times d}$ are the left and right singular vectors, and $\BB{\Lambda} \in \real{d \times d}$ is the diagonal matrix of singular values. As in the main text, we consider the case $d \leq n$. We recall the definitions of the projected state, observation noise,  and observation variables as $\breve{\State} = \BB{V}^\top \tilde{\State} \in \real{d}$, $\breve{\Noiseobs} = \BB{U}^\top \tilde{\Noiseobs} \in \real{d},$ and  $\breve{\Meas} = \BB{U}^\top \tilde{\Meas} \in \real{d}$. In the rotated spaces, the observation model can be written as:
\begin{equation}
\label{eqnSM:rotatedlik}
\breve{\Meas} = \BB{\Lambda} \breve{\State} + \breve{\Noiseobs},
\end{equation}
where the rotated observation operator $\BB{\Lambda}$ and the observation error covariance are diagonal matrices. Hence, the inference in this rotated space can be performed in a fully decoupled manner for each state variable. Using the decoupled observation model \eqref{eqnSM:rotatedlik}, the Kalman gain in the rotated space is given by: 
\begin{equation}
    \breve{\K} = \BB{\Lambda}(\BB{\Lambda}^2 + \id{d})^{-1}.
\end{equation}
Thus, the linear analysis map in the rotated space $\breve{\tmap}$ is given by:
\begin{equation}
\label{eqnSM:rotatedlinearmap}
    \breve{\tmap}(\breve{\meas}, \breve{\state}) = \breve{\state} - \breve{\K}(\breve{\meas} - \breve{\meas}^{\star}) =  \breve{\state} - \BB{\Lambda}(\BB{\Lambda}^2 + \id{d})^{-1}(\breve{\meas} - \breve{\meas}^{\star}),
\end{equation}
where $\breve{\meas}^{\star} = \BB{U}^\top \cov{\Noiseobs}^{-1/2} \meas^{\star}$ denotes the whitened and rotated realization of the assimilated data. The $d$ columns of $\BB{V}$ span the informative subspace of the whitened state space (\ie where the random variable $\tilde{\State}$ lives). Thus, $\proj_d = \BB{V} \BB{V}^\top \in \real{n \times n}$ is an orthogonal projector from the whitened state space to this informative subspace. Note that $\proj_d$ has rank $d$ as the subscript suggests. Therefore, the whitened state variable can be decomposed as:
\begin{equation}
\label{eqnSM:decomposition}
\begin{aligned}
    \tilde{\State} =  \proj_d \tilde{\State} + (\id{n} -  \proj_d) \tilde{\State} = \tilde{\State}^{\parallel} + \tilde{\State}^\perp,
\end{aligned}
\end{equation}
where $\tilde{\State}^{\parallel} \coloneqq  \proj_d \tilde{\State}$ and $\tilde{\State}^\perp \coloneqq (\id{n} -  \proj_d) \tilde{\State}$. 
Since $ \proj_d$ is an orthogonal projector, the decomposition above is unique and $\tilde{\State}^{\parallel} \in \real{n}$ is orthogonal to 
$\tilde{\State}^\perp \in \real{n}$. We can always complete the orthogonal family of columns $\BB{V}$ to form a basis for $\real{n}$ with $n-d$ orthogonal columns $\BB{V}^\perp$. The projector on this complementary subspace is given by $ \proj^\perp_{n-d} = \BB{V}^\perp {\BB{V}^\perp}^\top = \id{n} -  \proj_d$. We can now connect the decomposition~\eqref{eqnSM:decomposition} with the projected state variables on the informative and complementary subspaces: 
\begin{equation}
\begin{aligned}
\label{eqnSM:projectionstatetilde}
\tilde{\State} & = \BB{V} \BB{V}^\top \tilde{\State} + \BB{V}^\perp {\BB{V}^\perp}^\top \tilde{\State} = \BB{V} (\BB{V}^\top \tilde{\State}) + \BB{V}^\perp ({\BB{V}^\perp}^\top \tilde{\State}) = \BB{V} \breve{\State} + \BB{V}^\perp \State^\perp,
\end{aligned}
\end{equation}
with $\breve{\State} \in \real{d}$ and $\State^\perp \in \real{n-d}$. We emphasize that the inference performed with the analysis map in~\eqref{eqnSM:rotatedlinearmap} only acts on the rotated state variable $\breve{\State} \in \real{d}$, while the complementary component $\tilde{\State}^\perp$ of the whitened state is unaffected by the assimilation. Using eq.~\eqref{eqnSM:projectionstatetilde}, the low-rank analysis map in the original space is given by: 
\begin{equation} \label{eqnSM:analysismap_decomposed}
    \tmap(\meas, \state) =  \cov{\State}^{1/2}(\BB{V} \breve{\tmap}(\breve{\meas}, \breve{\state}) + (\id{d} - \BB{V}\BB{V}^\top) \cov{\State}^{-1/2}\state) = \state - \K(\meas - \meas^\star),
\end{equation}
where $\K$ denotes the Kalman gain. In the original space, we obtain the desired factorization of the Kalman gain $\K \in \real{n \times d}$ given by
\begin{equation}
\label{eqnSM:kalmangainlowrank}
    \K = \cov{\State}^{1/2}\BB{V}\BB{\Lambda}(\BB{\Lambda}^2 + \id{d})^{-1}\BB{U}^\top \cov{\Noiseobs}^{-1/2}.  
\end{equation}
In the whitened space (where $\tilde{\State}$ and $\tilde{\Meas}$ live), it is easy to see that $\BB{V}\BB{\Lambda}(\BB{\Lambda}^2 + \id{d})^{-1}\BB{U}^\top$ constitutes the singular value decomposition of the Kalman gain. Thus, in the whitened space, exploiting the spectrum of the whitened observation matrix gives us the best low-rank approximation of the Kalman gain. Unfortunately, \eqref{eqnSM:kalmangainlowrank} is no longer the SVD of the Kalman gain in the original space as the columns of the matrices $\cov{\State}^{1/2}\BB{V}$ and  $\cov{\Noiseobs}^{-\top/2}\BB{U}$ are not necessary orthogonal. Nonetheless, the proposed factorization gives us a constructive means to form a low-rank approximation of the Kalman gain in the original space (even if it is not optimal in the canonical Frobenius norm) without having to form the entire Kalman gain. This is clearly a desired feature for large inference problems where the Kalman gain cannot be practically formed and stored in memory.

\section{\label{secSM:nonlinearlowrank}Low-rank assimilation for the nonlinear--Gaussian case}

This section provides some intuition for  the construction of the state and observation Gramians in the nonlinear--Gaussian setting. In linear--Gaussian case, we show that the singular vectors of the whitened observation matrix (defined in Sec. \ref{secSM:linearlowrank}) are the directions maximizing Rayleigh quotients of interest. In the state space, it is enlightening to examine the directions that maximize the Rayleigh quotient of the posterior to prior precision in the linear--Gaussian case:
\begin{equation}
\label{eqnSM:Rpostprior} 
\mathcal{R}(\BB{w}) = \frac{\big \langle  \BB{w}, \cov{\State \given \Meas}^{-1} \BB{w} \rangle}{\langle \BB{w}, \cov{\State}^{-1}\BB{w} \rangle}, \mbox{ for } \BB{w} \in \real{n}.
\end{equation}
Given that $\cov{\State \given \Meas}^{-1} = \Obs^\top \cov{\Noiseobs}^{-1} \Obs + \cov{\State}^{-1}$, maximizing this Rayleigh quotient is equivalent to maximizing the Rayleigh ratio of the Hessian of the log-likelihood to the prior precision given by:
\begin{equation}
\label{eqnSM:Slinear} 
\mathcal{S}(\BB{w}) = \frac{\big \langle  \BB{w}, \Obs^\top \cov{\Noiseobs}^{-1} \Obs  \BB{w} \big \rangle}{\langle \BB{w}, \cov{\State}^{-1}\BB{w} \rangle}.
\end{equation}
This equation shows the connection between the directions that maximize the Rayleigh quotient of the posterior to prior, and the directions (in the state space) where the observations are most informative with respect to the prior.  With the change of variable $\BB{v} = \cov{\State}^{-1/2}\BB{w} \in \real{n}$, we also obtain
\begin{equation}
\label{eqnSM:Slinearwhitened} 
\tilde{\mathcal{S}}(\BB{v}) = \frac{\big \langle \BB{v}, \cov{\State}^{1/2}\Obs^\top \cov{\Noiseobs}^{-1} \Obs \cov{\State}^{1/2} \BB{v} \big \rangle}{\langle \BB{v} , \BB{v} \rangle}.
\end{equation}
 We denote the columns of $\V$ that span the image space of the projector $ \proj_{d}$ introduced in the previous section as $\{ \BB{v}_1, \ldots, \BB{v}_d \}$. Spantini et al.~\cite{spantini2015optimal} showed that the vector $\BB{v}_j$  maximizes the Rayleigh quotient $\tilde{\mathcal{S}}$ over the subspace $\real{n} \mbox{ \textbackslash \; span}\{\BB{v}_1, \ldots, \BB{v}_{j-1}\}$, which is the null space of the projector generated by the previous columns vectors $\{\BB{v}_1, \ldots, \BB{v}_{j-1}\}$. In other words, we are successively identifying the directions in the whitened state space where the observations are the most expressive relative to the prior and which are not in the same subspace as the previous directions. The matrix $\Cx = \cov{\State}^{1/2}\Obs^\top \cov{\Noiseobs}^{-1} \Obs \cov{\State}^{1/2} \in \real{n \times n}$ can in fact be rewritten as the inner product of the whitened observation matrix introduced in~\ref{secSM:linearlowrank}, \ie $\Cx = \tilde{\Obs}^\top \tilde{\Obs} $. It is straightforward to show that the vectors $\{ \BB{v}_1, \ldots, \BB{v}_d \}$ are the $d$ eigenvectors associated with the $d$ largest eigenvalues of the positive semi-definite matrix $\Cx$. The following $n-d$ eigenvectors form an orthonormal basis for the non-informative subspace $\BB{V}^\perp$. The interpretation of the vectors $\{ \BB{v}_1, \ldots, \BB{v}_n \}$ as the eigenvectors of the matrix $\Cx$ is an important step in generalizing to the setting with nonlinear observational models. In the nonlinear and non-Gaussian setting, Cui et al.~\cite{cui2021data} showed that the most important assimilation directions (for any realization of the observations) in the whitened state space can be identified by the eigenvectors of the \textit{state space Gramian}:
\begin{equation}
\label{eqnSM:cx}
    \Cx = \int \left(\cov{\Noiseobs}^{-1/2}\nabla \obs(\state) \cov{\State}^{1/2}\right)^\top \left(\cov{\Noiseobs}^{-1/2}\nabla \obs(\state) \cov{\State}^{1/2}\right)  \mathsf{d}\pdfprior(\state)\in \real{n\times n},
\end{equation}
where the expectation is taken over the prior distribution. As expected,  eq.~\eqref{eqnSM:cx} reverts to $\tilde{\Obs}^\top \tilde{\Obs}$ in the linear--Gaussian case. 

To the best of our knowledge, there is no proved procedure to identify the most important directions of the observation space for a nonlinear observation model. We propose a heuristic inspired from the construction of eq.~\eqref{eqnSM:cx} for the state space that reverts to the columns of the orthogonal matrix $\BB{U}$ in the linear--Gaussian case. It is reasonable to look for the directions in the observation space that maximize the relative ratio of the signal to the observational noise. In the linear--Gaussian case, we can form the Rayleigh quotient $\mathcal{T}$ that conveys this comparison as
\begin{equation}
\label{eqnSM:Tlinear} 
\mathcal{T}(\BB{q}) = \frac{\big \langle  \BB{q}, \cov{\Meas} \BB{q} \big \rangle}{\langle \BB{q}, \cov{\Noiseobs}\BB{q} \rangle} - 1 = \frac{\big \langle  \BB{q}, (\cov{\Meas} - \cov{\Noiseobs}) \BB{q} \big \rangle}{\langle \BB{q}, \cov{\Noiseobs}\BB{q} \rangle}, \mbox{ for } \BB{q} \in \real{d}.
\end{equation}
From eq.~\eqref{eqnSM:linobs}, we have $\cov{\Meas} = \Obs \cov{\State} \Obs^\top + \cov{\Noiseobs}$. With the change of variable $\BB{u} = \cov{\Noiseobs}^{1/2} \BB{q} \in \real{d}$, we obtain: 
\begin{equation}
\label{eqnSM:Tlinearwhitened} 
\tilde{\mathcal{T}}(\BB{u}) = \frac{\big \langle  \BB{u}, \cov{\Noiseobs}^{-1/2} \Obs \cov{\State} \Obs^\top \cov{\Noiseobs}^{-1/2} \BB{u} \big \rangle}{\langle \BB{u}, \BB{u} \rangle}
\end{equation}
It is easy to show that the directions that maximize the ratio  in~\eqref{eqnSM:Tlinearwhitened} are the eigenvectors of the matrix $\Cy = \cov{\Noiseobs}^{-1/2} \Obs \cov{\State} \Obs^\top \cov{\Noiseobs}^{-1/2} = \tilde{\Obs}\tilde{\Obs}^\top \in \real{d\times d}$. The eigenvectors of $\Cy$ are also the column vectors of the matrix $\BB{U}$ introduced in the previous section. Inspired by the treatment in the state space, we propose to use the eigenvectors of the \textit{observation space Gramian} $\Cy$ to select the important assimilation directions in the whitened observation space:
\begin{equation}
\label{eqnSM:cy}
    \Cy  = \int \left(\cov{\Noiseobs}^{-1/2}\nabla \obs(\state) \cov{\State}^{1/2}\right) \left(\cov{\Noiseobs}^{-1/2}\nabla \obs(\state) \cov{\State}^{1/2}\right)^\top  \mathsf{d}\pdfprior(\state) \in \real{d \times d}.
\end{equation}
Similarly to $\Cx$, eq.~\eqref{eqnSM:cy} reverts to $\tilde{\Obs} \tilde{\Obs}^\top$ in the linear--Gaussian case. 

\section{\label{secSM:algolowrank}Algorithm for the analysis step of the low-rank ensemble Kalman filter}

In this section, we present the complete algorithm for each iteration of the low-rank ensemble Kalman filter. The algorithm transforms a set of forecast samples to analysis samples after assimilating data $\meas^\star$.
\begin{algorithm}
\DontPrintSemicolon
    \KwInput{$\meas^\star \in \real{d}$, $\obs:\real{n} \to \real{d}$, $M$ samples $\{\state^i\}$ from $\pdfprior$, ranks $\rx$ and $\ry$}
    \KwOutput{$M$ samples $\{\state_a^i\}$ from $\pdfpost(\cdot \given \meas^\star)$}
    \vskip 0.2cm
    \tcc{Evaluate the observation operator for the prior samples, and sample from the noise distribution.}
    \For{$i= 1:M$}{
    $\var^i \leftarrow{} \obs(\state^i)$\\
    $\noiseobs^i \sim \N(\zero{d}, \cov{\Noiseobs})$
    }
    \tcc{Compute a Monte-Carlo approximation of $\Cx$ and $\Cy$.}
    $\sCx \leftarrow{} \frac{1}{M-1} \sum_{i=1}^M  \left(\cov{\Noiseobs}^{-1/2}\nabla \obs(\state^i) \cov{\State}^{1/2}\right)^\top \left(\cov{\Noiseobs}^{-1/2}\nabla \obs(\state^i) \cov{\State}^{1/2}\right) $\\
    $\sCy \leftarrow{} \frac{1}{M-1} \sum_{i=1}^M  \left(\cov{\Noiseobs}^{-1/2}\nabla \obs(\state^i) \cov{\State}^{1/2}\right) \left(\cov{\Noiseobs}^{-1/2}\nabla \obs(\state^i) \cov{\State}^{1/2}\right)^\top  $\\[2pt]
    \tcc{Perform a low-rank eigendecomposition of $\sCx$ and  $\sCy$.}
    $\BB{V}_{\rx} \xleftarrow{} \texttt{eigenvector}(\sCx, \rx)$\\
    $\BB{U}_{\rx} \xleftarrow{} \texttt{eigenvector}(\sCy, \ry)$\\[2pt]
    \tcc{Whiten and project the prior samples $\{\state^i\}$, observation samples $\{ \var^i \}$, and the observation noise samples $\{\noiseobs^i\}$, by substracting the empirical mean $\widehat{\BB{\mu}}$, and rotating the samples by $\BB{V}\cov{}^{-1/2}$.}
    \For{$i= 1:M$}{
    $\breve{\state}^{i} \xleftarrow{} \BB{V}_{\rx}^\top \cov{\State}^{-1/2}(\state^i - \widehat{\BB{\mu}}_{\State})$\\
    $\breve{\var}^{i} \xleftarrow{} \BB{U}_{\ry}^\top \cov{\Noiseobs}^{-1/2}(\var^i - \widehat{\BB{\mu}}_{\Var})$\\
    $\breve{\noiseobs}^{i} \xleftarrow{} \BB{U}_{\ry}^\top \cov{\Noiseobs}^{-1/2}(\noiseobs^i - \widehat{\BB{\mu}}_{\Noiseobs})$
    }
    \tcc{Form the perturbation matrices for the whitened state $\BB{A}_{\breve{\State}}\in \real{\rx \times M}$, the whitened observation $\BB{A}_{\breve{\Var}}\in \real{\ry \times M}$, and the whitened observation noise $\BB{A}_{\breve{\Noiseobs}}\in \real{\ry \times M}$.}
    \For{$i= 1:M$}{
    $\BB{A}_{\breve{\State}}[:,i] \xleftarrow{} \frac{1}{\sqrt{M-1}} \left(\breve{\state}^{i} - \widehat{\BB{\mu}}_{\breve{\State}} \right)$\\
    $\BB{A}_{\breve{\Var}}[:,i] \xleftarrow{} \frac{1}{\sqrt{M-1}} \left(\breve{\var}^{i} - \widehat{\BB{\mu}}_{\breve{\Var}} \right)$\\
    $\BB{A}_{\breve{\Noiseobs}}[:,i] \xleftarrow{} \frac{1}{\sqrt{M-1}} \left(\breve{\noiseobs}^{i} - \widehat{\BB{\mu}}_{\breve{\Noiseobs}} \right)$
    }
    \tcc{Apply the Kalman gain in the projected space based on the representers \cite{burgers1998analysis}.
    Solve the linear system for $\breve{\BB{b}} \in \real{\ry \times M}$:}
    
    $(\BB{A}_{\breve{\Var}}\BB{A}_{\breve{\Var}}^\top  + \BB{A}_{\breve{\Noiseobs}}\BB{A}_{\breve{\Noiseobs}}^\top)\breve{\BB{b}} =
    \BB{U}_{\ry}^\top\cov{\Noiseobs}^{-1/2}(\meas^\star \BB{1}_M^\top - (\BB{E}_{\Var} + \BB{E}_{\Noiseobs})) $,\\
    \tcc{where $\BB{E}_{\Var} \in \real{d \times M}$ and  $\BB{E}_{\Noiseobs} \in \real{d \times M}$ are the ensemble matrices of the samples $\{\var^i\}$ and $\{\noiseobs^i\}$. $\BB{1}_M$ denotes a vector of ones of length $M$.
    Lift result to the original space}
    \For{$i= 1:M$}{
    $\state^{i}_a \xleftarrow{} \state^{i} + \cov{\State}^{1/2} \BB{V}_{\rx}\left(\BB{A}_{\breve{\State}}\BB{A}_{\breve{\Var}}^\top\right)\breve{\BB{b}}[:,i]$
    }
    \vskip 0.2cm
    return $\{\state_a^i\}$
\caption{\texttt{lowrankenkf}$(\meas^\star, \obs, \{\state^i \}, \rx, \ry)$ assimilates the data $\meas^\star$ in the prior samples $\{\state^1, \ldots, \state^M\}$ with the low-rank EnKF. The integers $\rx$ and $\ry$ determine the rank of the projected subspace in the state space and in the observation space, respectively.}
\label{algo:lowrankenkf}
\end{algorithm}

\end{appendices}


\bibliographystyle{RS}
\bibliography{mybib.bib} 







\end{document}